\documentclass[11pt,a4paper]{article}
\pdfoutput=1
\usepackage{jcappub}
\usepackage{rotating}
\usepackage{array}
\usepackage{amsmath}
\usepackage[normalem]{ulem}
\usepackage{slashed}
\usepackage{booktabs}
\usepackage[pdftex,table]{xcolor}
\usepackage{units}
\usepackage{xspace}
\usepackage{makecell}

\newcolumntype{C}[1]{>{\centering\let\newline\\\arraybackslash\hspace{0pt}}m{#1}}

\newcommand{\ba}[1]{\ensuremath{\left( #1 \right)}}
\newcommand{\bb}[1]{\ensuremath{\left[ #1 \right]}}
\newcommand{\bc}[1]{\ensuremath{\left\{ #1 \right\}}}

\newcommand{\pd}[2]{\ensuremath{\frac{\partial #1}{\partial #2}}}

\newcommand{\nocontentsline}[3]{}
\newcommand{\tocless}[2]{\bgroup\let\addcontentsline=\nocontentsline#1{#2}\egroup}

\newcommand*\diff{\mathop{}\!\mathrm{d}}

\newcommand{\vw}{\ensuremath{v_w}\xspace}

\newcommand{\tot}{\ensuremath{\text{tot}}}
\newcommand{\DS}{\ensuremath{\text{DS}}}
\newcommand{\SM}{\ensuremath{\text{SM}}}
\newcommand{\perc}{\ensuremath{\text{p}}}
\newcommand{\peak}{\ensuremath{\text{peak}}}
\newcommand{\bro}{\ensuremath{\text{br}}}
\newcommand{\sym}{\ensuremath{\text{sym}}}
\newcommand{\nuc}{\ensuremath{\text{n}}}
\newcommand{\crit}{\ensuremath{\text{c}}}
\newcommand{\sw}{\ensuremath{\text{sw}}}

\newcommand{\EWPT}{\ensuremath{\text{EWPT}}}

\newcommand{\DM}{\ensuremath{\text{DM}}}
\newcommand{\GW}{\ensuremath{\text{GW}}}
\newcommand{\rad}{\ensuremath{\text{rad}}}
\newcommand{\true}{\ensuremath{\text{t}}}

\newcommand{\TDSbro}{\ensuremath{T_{\DS}^{\bro}}}
\newcommand{\TDSsym}{\ensuremath{T_{\DS}^{\sym}}}

\newcommand{\dd}{\mathrm{d}}

\newcommand{\ds}{{\sf DarkSUSY}}

\usepackage{multirow}
\usepackage{caption}
\usepackage{subcaption}

\arxivnumber{DESY-23-184, P3H-086, TTP-055}

\title{Hunting WIMPs with LISA: Correlating dark matter and gravitational wave signals}

\author[a]{Torsten Bringmann,}
\author[b]{Tom\'{a}s E. Gonzalo,}
\author[b]{Felix Kahlhoefer,}
\author[b,c]{Jonas Matuszak,}
\author[d]{and Carlo Tasillo}

\affiliation[a]{Department of Physics, University of Oslo, Box 1048, N-0316 Oslo, Norway}
\affiliation[b]{Institute for Theoretical Particle Physics (TTP), Karlsruhe Institute of
  Technology (KIT), 76128 Karlsruhe, Germany}
\affiliation[c]{Institute for Theoretical
  Particle Physics and Cosmology (TTK), RWTH Aachen University, D-52056 Aachen, Germany}
\affiliation[d]{Deutsches Elektronen-Synchrotron DESY, Notkestr.~85, 22607 Hamburg, Germany}

\emailAdd{torsten.bringmann@fys.uio.no, tomas.gonzalo@kit.edu, kahlhoefer@kit.edu, jonas.matuszak@kit.edu, carlo.tasillo@desy.de}

\abstract{The thermal freeze-out mechanism
  in its classical form is tightly connected to physics beyond the Standard Model around
  the electroweak scale, which has been the target of enormous experimental efforts. In
  this work we study a dark matter model in which freeze-out is triggered by a strong
  first-order phase transition in a dark sector, and show that this phase transition must
  also happen close to the electroweak scale, i.e.~in the temperature range relevant for
  gravitational wave searches with the LISA mission. Specifically, we consider the
  spontaneous breaking of a $U(1)^\prime$ gauge symmetry through the vacuum expectation value
  of a scalar field, which generates the mass of a fermionic dark matter candidate that
  subsequently annihilates into dark Higgs and gauge bosons. In this set-up the peak
  frequency of the gravitational wave background is tightly correlated with the dark
  matter relic abundance, and imposing the observed value for the latter implies that the
  former must lie in the milli-Hertz range. A peculiar feature of our set-up is that the
  dark sector is not necessarily in thermal equilibrium with the Standard Model during the
  phase transition, and hence the temperatures of the two sectors evolve independently.
  Nevertheless, the requirement that the universe does not enter an extended period of
  matter domination after the phase transition, which would strongly dilute any
  gravitational wave signal, places a lower bound on the portal coupling that governs the
  entropy transfer between the two sectors. As a result, the predictions for the peak
  frequency of gravitational waves in the LISA band are robust, while the amplitude can
  change depending on the initial dark sector temperature.}

\keywords{primordial gravitational waves (theory), cosmology of theories beyond the SM,
  particle physics -- cosmology connection}

\begin{document}
\maketitle
\flushbottom

\section{Introduction}
\label{sec:introduction}

Dark matter (DM) is known to be the dominant form of matter in the universe, but it has so
far evaded any attempt of detection in the laboratory or by other non-gravitational
means~\cite{ParticleDataGroup:2022pth}. These null results have cast doubt on the
so-called WIMP miracle, where DM is produced from the thermal bath of Standard Model (SM)
particles in the early universe, and which for a long time has been used to motivate
sizeable couplings between DM particles and the SM. Indeed, it has been shown
that thermal freeze-out may happen entirely within an extended dark sector, such that the
observed DM relic abundance $\Omega_\text{DM} h^2 \simeq 0.12$~\cite{Planck2018} can be reproduced without the
need for any sizeable couplings between the dark and the visible
sector~\cite{Pospelov:2007mp}. These so-called secluded DM models pose a great challenge
for laboratory searches due to their apparent lack of testable predictions.

At the same time, gravitational wave (GW) observatories have opened a completely new window
into the universe, making it possible to observe objects and phenomena that affect visible
matter only through gravity. The
proposed LISA mission~\cite{LISA:2017pwj} will extend this window to the mHz frequency range, allowing in
particular for the observation of a stochastic GW background that would be
connected to a strong first-order phase transition (PT) close to the electroweak
scale~\cite{Caprini2015, Caprini:2018mtu, Caprini2019}. LISA therefore raises new hopes to
detect dark sectors that are otherwise unobservable.
Over the past few years, first-order PTs in dark sectors have been studied in great
detail~\cite{Espinosa:2008kw, Schwaller:2015tja, Breitbach2019, Ertas:2021xeh,
  Bringmann:2023opz}, and various correlations between GW signals and the phenomenology of
DM have been explored~\cite{Alanne:2014bra, Baldes:2017rcu, Hall:2019rld, Baker2019,
  Azatov2021, Baker:2021nyl, Kierkla:2022odc, Lewicki:2023mik, Lewicki:2023ioy,
  Kanemura:2023jiw, Biondini:2022ggt, Arcadi:2023lwc,madge2019}. The conclusion of these
studies is that it is difficult to robustly predict the expected amplitude of the GW
signal for a given DM model, because strong PTs often only happen in special regions of
parameter space. In other words, it appears generally challenging to identify a strong
correlation between the GW {\it amplitude} and the DM abundance. In this work, we instead
focus on the {\it peak frequency} of the GW signal and show that it can be tightly
correlated with the predicted DM relic abundance. Intriguingly, when imposing the observed
value of $\Omega_\text{DM} h^2 = 0.12$ and focussing on GW signals strong enough to be
potentially observable, we predict a GW peak frequency that falls right into the most
sensitive range of LISA.

Before describing our analysis in detail, let us provide a rough sketch of the argument.
We consider a dark sector comprised of a fermionic DM candidate $\chi$ charged under a new
$U(1)'$ gauge group that is spontaneously broken by the vacuum expectation value (vev)
$v_\phi$ of a new dark Higgs field. It is well known that strong PTs can occur in this model for a sufficiently large gauge coupling~\cite{Croon:2018kqn,Athron:2023aqe}. All newly introduced particles are massless before symmetry breaking and
acquire a mass proportional to $v_\phi$ afterwards. The dark gauge boson $A'$ (a.k.a.~dark
photon) and the dark Higgs boson $\phi$ are generally unstable against decays into SM
particles, but $\chi$ is stable and may obtain a sizeable relic abundance through thermal
freeze-out. If the spontaneous symmetry breaking occurs in a first-order PT,
bubbles of the new phase will nucleate spontaneously, expand and collide. This process
perturbs the dark plasma and leads to the emission of GWs, with a present-day peak frequency very roughly given by~\cite{Caprini2019}
\begin{align}
\label{eq:fpeak_simp}
  f_{\rm peak} \simeq 10 \, \text{mHz} \ba{\frac{\beta/H}{100}} \ba{\frac{T_\perc}{1 \, \text{TeV}}}
  .
\end{align}
Here $\beta/H$ denotes the speed of the PT and $T_\perc$ is the temperature of the SM
heat bath at the time of percolation. For a not-too-strongly supercooled dark sector PT, which is what we
consider here, one expects $\beta / H \sim 100$ and $T_\perc \sim v_\phi$.

The relic density from thermal freeze-out, on the other hand,  can in leading-order approximation
be written as~\cite{Kolb:1990vq}
\begin{align}
  \Omega_\text{DM} \simeq 0.1 \,
  \frac{10^{-8} \, \text{GeV}^{-2}}{\langle \sigma_\text{ann} v \rangle} \,,
\end{align}
with ${\langle \sigma_\text{ann} v \rangle}$ the thermally averaged DM annihilation cross section.
If the DM particles dominantly annihilate into the dark Higgs bosons $\phi$, arising from the same dark
Higgs field that generates the DM mass, it is parametrically of the form
\begin{equation}
 \langle \sigma_\text{ann} v \rangle \sim \frac{y^4}{m_\text{DM}^2} \sim \frac{y^2}{v_\phi^2} \,,
\end{equation}
where $y$ denotes the DM Yukawa coupling. At first sight, this coupling is arbitrary, and hence the freeze-out mechanism
does not predict a specific dark sector mass scale. However, if we are interested in dark sectors that produce strong
first-order PTs and large GW signals, the dark gauge coupling $g$ and the dark Higgs quartic coupling $\lambda$ must be sizeable,
which implies that the dark Higgs boson mass $m_\phi$ cannot be much smaller than $v_\phi$. At the same time,
 the observed DM relic abundance can only be obtained through dark sector freeze-out
 if the DM particle is not the lightest particle in the dark sector
 (or at least not much lighter than its annihilation products~\cite{DAgnolo:2015ujb}).
This, in turn, implies that $y$ cannot be much smaller than unity, and hence $v_\phi \sim \mathrm{TeV}$ once we require
$\Omega_\text{DM} h^2 \sim 0.1$.
Combining this with the conclusion from eq.~\eqref{eq:fpeak_simp}, we thus expect a peak frequency
of $f_{\rm peak} \sim 10 \, \text{mHz}$ --  which, as advocated, lies right within the LISA band.

A possible concern with the simplified reasoning above is that a large Yukawa coupling
will affect the effective potential and may possibly prevent a first-order phase
transition, or even destabilize the scalar potential~\cite{Weinberg:1976pe}. We have also neglected the impact of additional DM
annihilation channels involving dark photons. In our full analysis, we explore the entire
parameter space of the model, calculating in detail the effective potential, the
thermodynamic quantities characterising the PT and the relic density from
thermal freeze-out. We then identify viable combinations of the different dark sector
couplings and show that the qualitative argument from above is confirmed by quantitative
calculations. In order to further refine the analysis, we also perform parameter scans
over all relevant model parameters -- namely the three couplings $g$, $\lambda$ and $y$ and the
dark Higgs vev $v_\phi$, and we identify parameter points for which the correct DM abundance is
obtained. Interpreting the sampling distributions for the model parameters as prior
probabilities thus enables us to define ``typical'' model predictions and quantify the
probability (in the Bayesian sense) of a detectable signal.

A significant focus of our analysis
is to extend the simple argument sketched above to situations where the couplings are so weak that the dark and
visible sectors do not necessarily share a common temperature, which would be maintained through (inverse) decays of SM
and dark Higgs bosons. Indeed, even if the two sectors have the same temperature initially, the first-order PT in the dark sector will change the temperature ratio, as the vacuum energy in the dark Higgs field is converted to rest mass and kinetic energy. This additional energy needs to be rapidly transferred to the SM in order to avoid a dilution of GW signals from late-time entropy injection~\cite{Ertas:2021xeh,Bringmann:2023opz}. We calculate the dilution of the GW background and derive a lower bound on the portal coupling from the requirement that no significant dilution occurs. We show that the portal coupling required for this purpose is well below the sensitivity of laboratory experiments.
Finally, we explore what happens if the initial temperature ratio of the two sectors
differs from unity. In this case the amplitude of the GW signal will
change~\cite{Breitbach2019,Fairbairn:2019xog,Li:2023bxy} -- but the peak frequency remains almost
unaffected, such that the estimate from above remains robust even for portal couplings
that are too small to quickly (re-)thermalize the sectors after the transition. This
conclusion is only modified if the portal coupling is so weak that the energy density of
the dark sector cannot be depleted and starts to dominate the energy density of the
universe.

The remainder of this work is structured as follows. In section~\ref{sec:model} we
introduce the model under consideration and discuss the
finite-temperature effective potential. We also briefly review the calculation of the
temperature and strength of the PT, and we identify the interesting regions of
parameter space. In section~\ref{sec:relic} we calculate the DM
relic density under the assumption that the dark and SM sector remain in thermal
equilibrium throughout their evolution, and explore the correlation between the relic
density and the GW signal. We revisit this assumption in section~\ref{sec:equilibrium}, and discuss in detail the
processes that thermalize the dark sector with itself and with the SM.
In section~\ref{sec:hot} we finally calculate the effect of
inefficient thermalisation on the GW signal. We consider the dilution due to entropy injection and show that for
hot dark sectors a net enhancement of the GW amplitude can remain, while the peak frequency is essentially unaffected. We
conclude in section~\ref{sec:conclusions} with a summary of our results and some
remarks about their consequences. In two technical appendices, we provide details on the bubble
wall velocity (appendix~\ref{app:darkwalls}) and on the Boltzmann equations for entropy transfer
(appendix~\ref{app:thermalisation}).

\section{Dark sector phase transition}
\label{sec:model}

\subsection{Dark sector model}

The model we study in this work is an extension of the models considered in refs.~\cite{holdom1986,Ertas:2021xeh}
and consists of a complex scalar $\Phi$ charged under the $U(1)'$ gauge symmetry, the
associated gauge boson $A'_\mu$, and two chiral fermion DM candidates, $\chi_L$ and
$\chi_R$. The Lagrangian describing the model is
\begin{align}
  \mathcal{L} = & |D_\mu \Phi|^2 - \frac{1}{4} A'_{\mu\nu} A^{'\mu\nu} + \mu^2 \Phi^*\Phi - \lambda(\Phi^*\Phi)^2 \notag \\
    &+ \chi_L^\dagger i \slashed{D}\chi_L + \chi_R^\dagger i \slashed{D}\chi_R - y\Phi \chi_L^\dagger\chi_R - y\Phi^*\chi_R^\dagger \chi_L\,,
      \label{Lagrangian1}
\end{align}
where $A'_{\mu\nu}$ is the field strength tensor of $A'_\mu$, $\lambda$ and $y$ are dimensionless
couplings and $\mu$ is a bare mass parameter for $\Phi$. The complex scalar and fermions are
charged under the $U(1)'$ group as $Q_\Phi = +1$, $Q_{\chi_L} = +1/2$ and $Q_{\chi_R} = -1/2$.
The kinetic mixing of the dark photon with SM hypercharge and the portal coupling of the scalar field $\Phi$ with the SM Higgs
field are assumed to be small enough that they satisfy experimental bounds (see, e.g., refs.~\cite{Ilten:2018crw,Bauer:2018onh}
for bounds on dark photons and ref.~\cite{Ferber:2023iso} for a recent review of bounds on Higgs mixing) and do not play a role
during the PT. We will return to these terms in section~\ref{sec:thermalization} when discussing the thermalisation of the dark
sector with the SM bath.

The tree-level scalar potential of our model has a minimum at $v_\phi = \pm \sqrt{\mu^2/\lambda}$. One can hence
expand the complex field as
$\Phi = (v_\phi + \phi + i\varphi)/\sqrt{2}$, where $\phi$ and $\varphi$ are real scalar fields. In addition, the
chiral fermions $\chi_L$ and $\chi_R$ can be written as a Dirac fermion $\chi$. The
Lagragian in eq.~\eqref{Lagrangian1} can thus be re-written as
\begin{align}
  \mathcal{L} = & \frac{1}{2}\partial_\mu \phi \partial^\mu \phi + \frac{1}{2} \partial_\mu\varphi \partial^\mu\varphi - \frac{1}{4} A'_{\mu\nu}A^{'\mu\nu}
       - \frac{1}{2} m_\phi^2\phi^2 + \frac{1}{2}m_{A'}^2A^{'2}_\mu  \notag \\
    &-gA'_\mu[\varphi\partial^\mu\phi - \phi\partial^\mu\varphi - v_{\phi}\partial^\mu\varphi] + \frac{g^2}{2}\phi^2A^{'2}_\mu + \frac{g^2}{2}\varphi^2 A^{'2}_\mu + g^2v_{\phi}\phi A^{'2}_\mu \notag \\
    &-\lambda v_{\phi}\phi^3 - \lambda v_{\phi}\varphi^2\phi - \frac{\lambda}{4}\phi^2\varphi^2 - \frac{\lambda}{4}\phi^4 -\frac{\lambda}{4} \varphi^4\notag\\
      &+i\bar\chi\slashed{\partial}\chi - m_\chi\bar\chi\chi + \frac{g}{2}\bar\chi\slashed{A}'\gamma^5\chi -
        \frac{y}{\sqrt{2}}\phi\bar\chi\chi + i\frac{y}{\sqrt{2}}\varphi\bar\chi\gamma^5\chi  \,,
      \label{eq:Lagrangian2}
\end{align}
where $g$ is the gauge coupling associated with the $U(1)'$ symmetry, and the bare masses
of the various fields depend on  the vacuum expectation value (vev) $v_\phi$ as
\begin{equation}
  m_\phi^2 = -\mu^2 + 3\lambda v_\phi^2 = 2\lambda v_\phi^2, \quad m_\varphi^2 = 0, \quad m_{A'}^2 = g^2 v_\phi^2, \quad m_\chi^2 = \frac{y^2}{2}v_\phi^2.
  \label{masses}
\end{equation}

\subsection{The effective potential}
\label{sec:Veff}

The properties of the PT can be computed from the temperature-dependent effective
potential $V_{\rm eff}$. This effective potential can be constructed out of the tree-level
potential extracted from eq.~\eqref{eq:Lagrangian2}, the 1-loop Coleman-Weinberg (CW)
potential $V_{\rm CW}$ and the finite temperature potential $V_T$~\cite{Quiros:1999jp}. The
CW potential in our model, computed using the background field ($\phi_b$)
method~\cite{Coleman:1973jx}, in the Landau gauge and the $\overline{\text{MS}}$
renormalisation scheme, has the form~\cite{Basler:2018cwe,Athron:2023xlk}
\begin{equation}
  V_{\rm CW}(\phi_b) = \sum_{a=\phi,\varphi,A',\chi} \pm g_a \frac{m_a^4(\phi_b)}{64\pi^2}
  \left(\log\frac{m_a^2(\phi_b)}{\Lambda^2} - k_a\right),
  \label{CWpotential}
\end{equation}
where the $+$ ($-$) sign applies to bosons  (fermions), $g_a$ are the degrees of
freedom of particle $a$, $m_a(\phi_b)$ are the background field-dependent masses, and
$k_a = 3/2$ for scalars and fermions and $k_a = 5/6$ for gauge bosons. In this study we
take the renormalisation scale $\Lambda$
to be the vev at zero temperature,
$v_\phi(T=0)$. The field-dependent masses in the CW potential are computed with
renormalised parameters, i.e.~using the expressions from eq.~\eqref{masses} with the
replacement $v_\phi \to \phi_b$ and with the addition of relevant counterterms~\cite{Quiros:1999jp}.
We choose the renormalisation conditions such that the dark Higgs mass and vev are fixed
to their tree-level values at zero temperature.

\begin{figure}[t]
 \centering
 \includegraphics[width=0.65\textwidth]{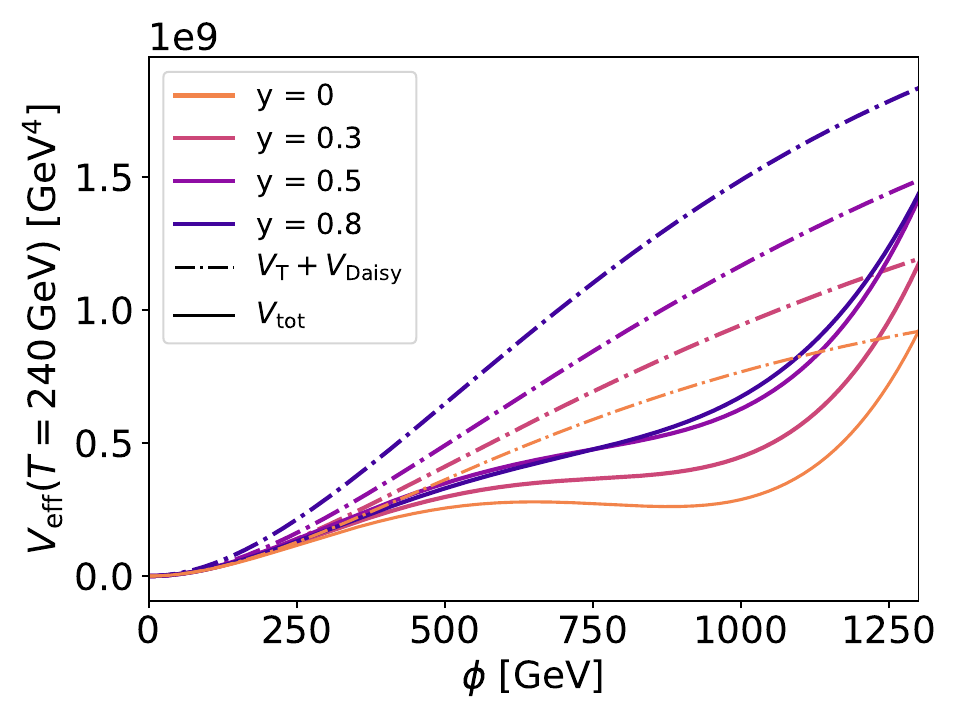}
 \caption{Total effective potential (solid) and finite temperature part (dot-dashed)
  for $g=0.67, \lambda=0.0035, v_{\phi} = 1000\,$GeV and $T = 240\,$GeV, for varying
   values of $y$.}
 \label{fig:potential}
\end{figure}

The finite-temperature part has the form
\begin{equation}
  V_T(\phi_b) = \frac{T^4}{2\pi^2} \sum_a \pm g_a J_{\text{b}/\text{f}}\left(\frac{m_a^2(\phi_b)}{T^2}\right),
  \label{finiteT}
\end{equation}
with the thermal functions $J_{\text{b}/\text{f}}$ for bosons/fermions defined in
ref.~\cite{Quiros:1999jp}. The finite temperature potential in eq.~\eqref{finiteT} suffers
from infrared divergences for bosonic modes when $T \gg m$, leading to a breakdown of
finite-temperature perturbation theory at high temperatures~\cite{Athron:2023xlk}. The
resummation of these modes is commonly done by ``daisy
resummation''~\cite{Carrington1992}. Here we follow the Espinosa method~\cite{arnold1994},
which adds an additional contribution to the effective potential of the form
\begin{equation}
  V_{\rm daisy}(\phi_b) = -\frac{T}{12\pi}\sum_{a=\phi,\varphi,A_\text{L}'} g_a \left[(m_a^2(\phi_b) + \Pi_a(T))^{3/2} - (m_a^2(\phi_b))^{3/2}\right],
\end{equation}
where $a$ now only runs over the scalar fields and the longitudinal component of the dark photon.
The one-loop thermal masses $\Pi_a$ in our model are given by
\begin{align}
  \Pi_\phi &= \left(\frac{\lambda}{3} + \frac{y^2}{12} + \frac{g^2}{4}\right)T^2 \,,\notag \\
  \Pi_\varphi &= \left(\frac{\lambda}{3} + \frac{y^2}{12}+\frac{g^2}{4}\right)T^2 \,,\notag \\
  \Pi_{A'} &= \left(\frac{1}{12} + \frac{2}{3}\right)g^2T^2 \,.
\end{align}
We only resum the Matsubara zero modes in the daisy potential, which are the ones causing the infrared divergences of eq.~\eqref{finiteT}, such that the thermal mass of the DM fermion does not enter our calculations. The final form of the effective potential is then given by
\begin{equation}
 V_{\rm eff}(\phi_b, T) = V_0(\phi_b) + V_{\rm CW}(\phi_b) + V_T(\phi_b, T) + V_{\rm daisy}(\phi_b, T)\, .\label{eq:effectivepotential}
\end{equation}
For illustration, we plot this potential in figure~\ref{fig:potential} for a choice of parameters
$g=0.67, \lambda=0.0035$, and varying values of $y$.

In addition to encoding the properties of the PT, the effective potential also provides
information about the stability of the true vacuum after the phase transition occurs.
In fact, a new
feature becoming important for non-zero Yukawa couplings is that for low values of $\lambda$
and $g$ the potential can become unbounded from below~\cite{Weinberg:1976pe}.
To ensure vacuum stability we require that no deeper vacua are present at zero
temperature. The requirement of a dark sector PT already implies that
$V_\text{eff}(0) > V_\text{eff}(v_\phi)$. Hence, it is sufficient to check whether there exist
vacua with lower potential energy for large field values, i.e.\ whether
$V_{\rm eff}(\phi_b) < V_{\rm eff}(v_\phi)$ for $\phi_b \gg v_\phi$.
In our analysis,
we explicitly exclude such parameter points.

It is well known that the one-loop, daisy-resummed calculation of the effective potential
can suffer from large theoretical uncertainties, foremost sourced by a large
renomalization scale-dependence~\cite{Croon:2020cgk}. A possibility to improve upon those
uncertainties is to systematically resum higher orders of the thermal masses in the
effective field theory framework of dimensional reduction~\cite{GINSPARG1980388}. In order
to validate our simpler approach, we therefore also implemented our model in \texttt{DRalgo}
\cite{Ekstedt:2022bff}, which automates the task of dimensional reduction. We calculate
the critical temperature in both our four-dimensional implementation and the reduced
three-dimensional theory for the parameter space where we expect a first-order PT. In the
regime where the effective field theory is valid ($T \gg m_{\phi}$) we find that the two results agree very
well. We therefore conclude that we can take the computationally more economical approach
of using the 1-loop, daisy-resummed effective potential stated in eq.~\eqref{eq:effectivepotential}.

\subsection{Properties of the phase transition}
\label{sec:PT}

The PT in our model occurs when the dark Higgs $\phi$ acquires a
non-vanishing vev at the minimum of the effective potential $V_{\rm eff}$, thereby
breaking the $U(1)'$ symmetry. At the nucleation temperature $T_\nuc$ the transition from the unbroken (false) vacuum to the
broken (true) vacuum becomes energetically favourable and bubbles of the new phase start
nucleating and expand rapidly to fill the entire Universe. The probability of
transitioning from one phase to the other, parametrized by the bubble nucleation rate
$\Gamma(T)$, can be computed
from the effective potential in a semiclassical formalism by solving the bounce
equation~\cite{Coleman1977,Linde1981,Linde1980}, which takes into account thermal
tunneling through the potential barrier. We use
\texttt{TransitionListener}~\cite{Ertas:2021xeh} for this purpose, an extension of \texttt{CosmoTransitions}
\cite{Wainwright:2011kj}, which takes care of the computation of the bounce action, as
well as phase tracing and the calculation of the thermodynamic properties of the phase
transition. For the moment, we will assume equal temperatures for the SM bath and the dark
sector, i.e.~we adopt a temperature ratio of 
\begin{equation}
\xi \equiv \frac{T_\DS}{T} =1 \, .
\end{equation}
The temperature that appears
in the effective potential in eq.~\eqref{eq:effectivepotential} and in the thermodynamic
quantities discussed below can thus be identified with the temperature of SM photons. We
discuss the general case with $\xi \neq 1$ in sections~\ref{sec:equilibrium} and \ref{sec:hot}.

Gravitational waves from a PT are produced when the expanding bubbles or their sound
shells collide (more in section \ref{sec:GWspec}). Therefore, the thermodynamic
properties of the PT must be computed at a time when the rate of bubble collisions
is maximised. This occurs at the time of \textit{percolation}, when the Universe is
permeated with a connected web of bubbles of the broken phase~\cite{guo2021,
  Athron:2023rfq, Caprini2019}, which happens when approximately 70\,\% of the Universe is
in the symmetric phase.
Quantitatively, the fraction of the Universe remaining in the
false vacuum after the PT is given by $P(T) = \exp \bb{-I(T)}$ with~\cite{Athron:2023xlk}
\begin{align}
	I(T) = \frac{4 \pi}{3} \vw^3 \int^{T_\crit}_{T} \diff T^\prime \, \frac{\Gamma(T^\prime)}{T^{\prime\,4} \, H(T^\prime) } \ba{\int_{T}^{T^\prime}  \frac{\diff T^{\prime\prime}}{H(T^{\prime\prime})}}^3 , \label{eq:falsvacfrac}
\end{align}
where $T$ is the common temperature of the dark sector (DS) and the SM bath, $\Gamma(T)$ is the
bubble nucleation rate, $H(T)$ is the Hubble parameter, $\vw$ is the wall velocity and
$T_\crit$ is the
`critical' temperature where the minimum of
the effective potential with non-vanishing vev becomes a global
minimum~\cite{PhysRevD.23.876}. Hence, the temperature at which percolation occurs can be
computed by solving $I(T_\perc) = 0.34$~\cite{Ellis:2018mja} with the approximation $\vw \rightarrow 1$.
A discussion about the bubble wall velocity in our model can
be found in appendix~\ref{app:darkwalls}.

\begin{figure}[t]
  \centering
  \includegraphics[width=.95\textwidth]{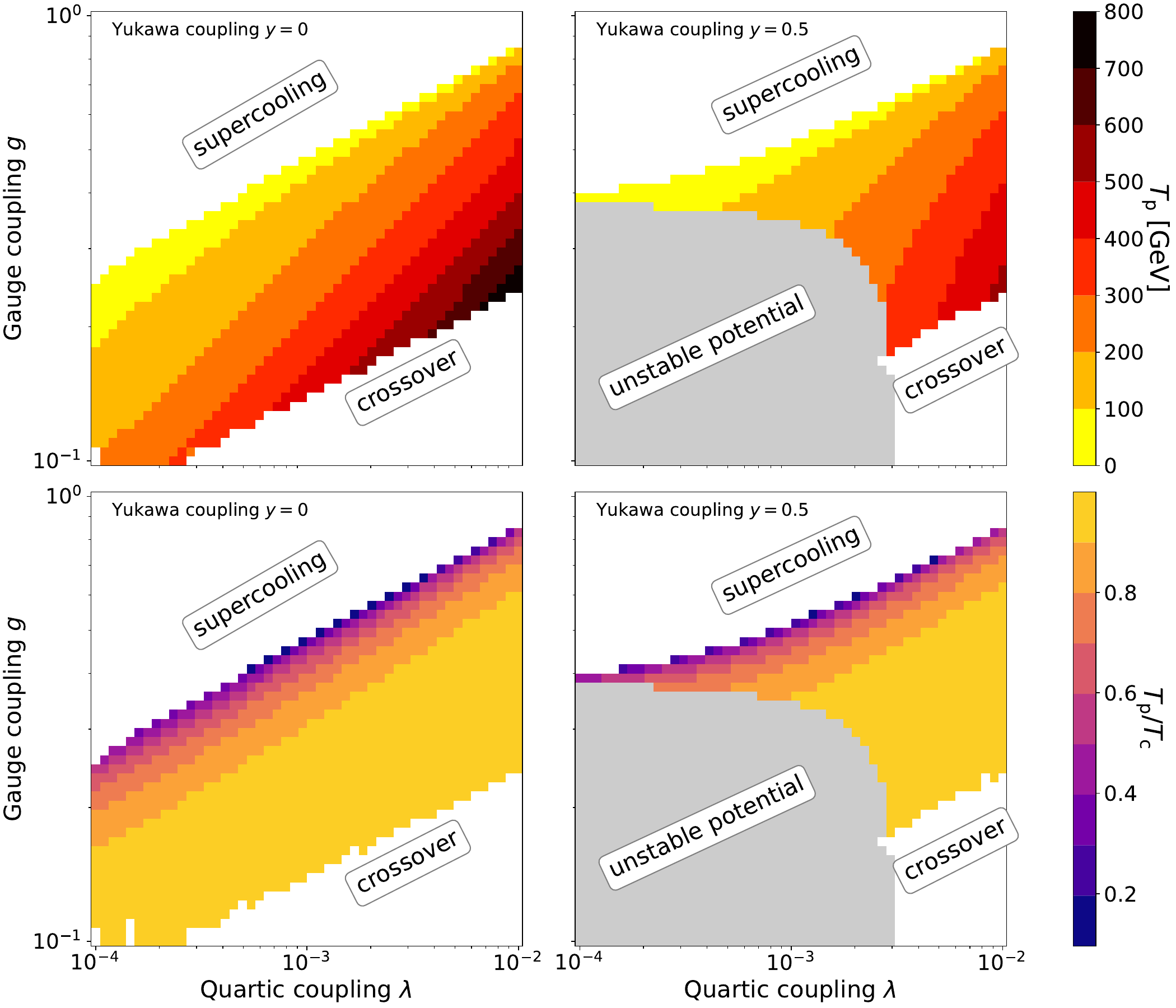}
  \caption{The percolation temperature $T_\perc$ (top) and the ratio of percolation
    temperature and critical temperature $T_\perc/T_\crit$ (bottom) in the $\lambda-g$ plane for
    Yukawa couplings of $y=0.0$ (left) and $y=0.5$ (right) and a vev of $v_\phi = 1$\,TeV.
    The coloured band shows the parameter region where a first-order PT is possible.}
  \label{fig:GWparams-T}
\end{figure}

The dependence of the percolation temperature $T_\perc$ on the model parameters can be seen in
figure~\ref{fig:GWparams-T}. In the top panels $T_\perc$ is shown as a function of the quartic
coupling $\lambda$ and gauge coupling $g$ for two values of the Yukawa coupling, $y = 0$ (left) and
$y = 0.5$ (right). There is a strong correlation between the values of $\lambda$ and $g$ that
produce a first-order PT, with lower values of $T_\perc$ in the top-left end of
the allowed band, and higher values of $T_\perc$ in the bottom-right. The disallowed areas
correspond to parameter regions where the transition is not first-order or it does not occur at all.
These effects are better illustrated in the bottom panels, where the color scale indicates the
ratio $T_\perc/T_\crit$ in the same parameter plane. The amount of supercooling of the transition is largest when $T_\perc$ is much lower than $T_ \crit$ and smallest when both temperatures almost coincide. For the points above the coloured contours, the potential barrier becomes so large that the bubble nucleation rate is too low for the transition to reach percolation; the region below instead indicates a smooth crossover transition in which no bubbles form since the potential does not develop a barrier between the phases. For non-zero values of the Yukawa coupling $y$, the enhanced thermal corrections
in the effective potential cause a delay of the development of the true vacuum
(cf.~figure~\ref{fig:potential}), thereby decreasing the value of $T_\perc$. The vacuum also
becomes deeper due to the Yukawa coupling, which increases the tunneling rate close to the
supercooled region, and thus slightly larger values of $g$ are within the allowed band.
The grey shaded regions, finally, indicate parameter combinations where the potential is unstable
(as discussed above in section \ref{sec:Veff}).

The strength of the PT is quantified by the amount of released latent heat, described by the difference in the trace 
$\theta \equiv g_{\mu \nu} T^{\mu \nu}$ of the energy momentum tensors between the two phases~\cite{Giese:2020rtr},
\begin{align}
	\Delta \theta = \bb{ - 4 \Delta V_\text{eff} + T \pd{\Delta V_\text{eff}}{T} }_{T_\perc}  > 0\, ,
	\label{theta}
\end{align}
where $\Delta V_\text{eff}$ is the difference in the effective potential between the vacua of the
two phases. A dimensionless quantity for the strength of the PT can be obtained by dividing $\Delta \theta$ by the total energy
density of the SM and dark sector plasma,
\begin{align}
	\alpha \equiv \frac{\Delta \theta / 4}{\rho_{\SM,\perc} + \rho_{\DS,\perc}} \,,
	\label{alpha}
\end{align}
where $\rho_{\SM,\perc}$ ($\rho_{\DS,\perc}$) is the energy density in the SM (DS) at percolation.
The last missing piece for obtaining a GW spectrum is the speed of the PT, which can be determined through
\begin{align}
  \beta/H = T_\perc \left. \frac{\diff}{\diff T}  \frac{S_3(T)}{T} \right|_{T=T_\perc} \, ,
\end{align}
where $S_3$ is the $O(3)$-symmetric bounce action for thermal tunneling.

\begin{figure}[t]
  \centering
  \includegraphics[width=.95\textwidth]{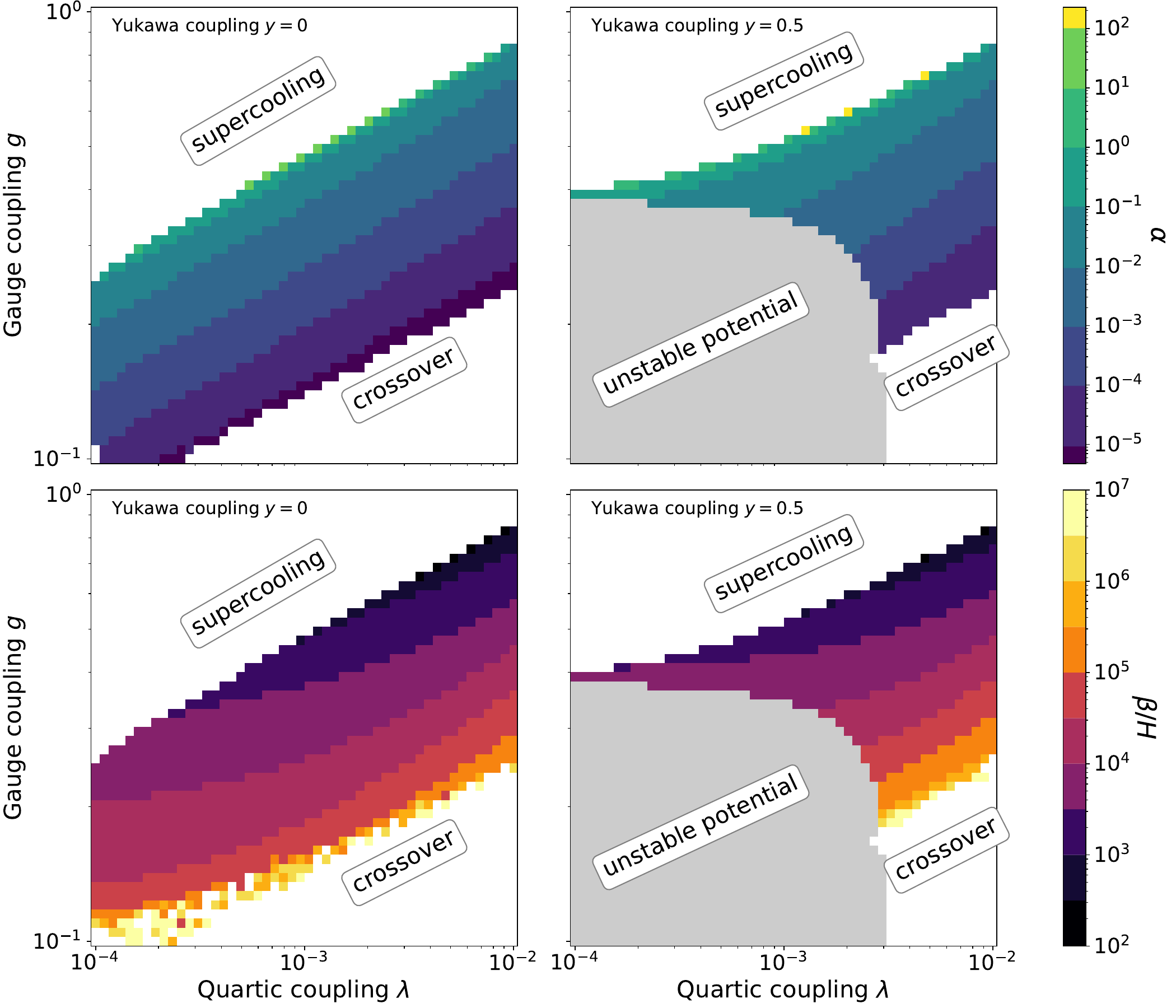}
  \caption{The transition strength $\alpha$ (top) and speed $\beta/H$ (bottom) of the PT, in the
    $\lambda-g$ plane for $y = 0$ (left) and $y = 0.5$ (right), with $v= 1$\,TeV and $\xi = 1$.}
  \label{fig:GWparams-alpha-beta}
\end{figure}

In figure~\ref{fig:GWparams-alpha-beta} we show how the transition strength $\alpha$ and transition speed $\beta/H$ depend on the model parameters, $\lambda$, $g$ and $y$.
The PT is relatively strong for most of the allowed region $\alpha \in (10^{-2}, 10^2)$ and it is
particularly strong close to the supercooled limit, where percolation is delayed
($T_\perc \ll T_\crit$). On the other hand, the speed of the PT $\beta/H$ becomes smaller in the
supercooling limit, reaching values of $\beta/H \approx 10^2 - 10^3$.

\subsection{Gravitational wave spectrum}
\label{sec:GWspec}

The spectrum of GWs in our scenario is produced dominantly through bulk fluid
motion in the reheated plasma due to the large velocity-dependent friction from the emission of
soft dark photons in the bubble wall, yielding a terminal bubble wall
velocity~\cite{Bodeker:2017cim,Gouttenoire:2021kjv, Azatov:2023xem}. A discussion of this argument can be found in
appendix~\ref{app:darkwalls}. As the case of
runaway bubbles can hence be excluded, we neglect the contribution of bubble collisions to the GW signal. Since the onset of
turbulence as a GW source is not yet understood well enough to make quantitative statements~\cite{Caprini2019}, and often
requires complicated lattice simulations~\cite{RoperPol:2019wvy}, we conservatively consider
sound waves the only relevant source of GWs emitted during the PT.
Therefore, the GW spectrum is exclusively
determined by the set of parameters $\bc{\alpha, \beta/H, T_\perc}$. We use semi-analytical
approximations to compute the peak frequency and spectrum of GWs from sound
shell collisions, based on simulations~\cite{Caprini2019}.
Our computation of the GW spectrum includes some
corrections taking into account that the transition happens in a dark sector and that the
bubble wall dynamics are independent of the SM field content. The spectrum of gravitational waves
is thus computed as~\cite{Bringmann:2023opz}
\begin{align}
	h^2 \Omega_\GW(f) &=
	\mathcal{R}h^2 \,  \tilde{\Omega}  \ba{\frac{\kappa_\sw \, \alpha}{\alpha + 1}}^2 \ba{\frac{\beta}{H}}^{-1}  \mathcal{Y} \,  S(f) \, .
	\label{spectrum}
\end{align}

In this expression, the prefactor $\mathcal{R}h^2$ quantifies the redshift of the amplitude of the emitted GW signal and is given
by~\cite{Ertas:2021xeh, Bringmann:2023opz}
\begin{align}
	\mathcal{R} h^2 &= \Omega_\gamma h^2  \left(\frac{h_{\SM,0}}{h_{\tot,\perc}}\right)^{4/3}  \left(\frac{g_{\tot,\perc}}{g_{\gamma,0}}\right) = 1.653 \cdot 10^{-5} \ba{\frac{100}{h_{\tot,\perc}}}^{4/3} \ba{\frac{g_{\tot,\perc}}{100}}  \,, \label{redshift}
\end{align}
where $\Omega_\gamma h^2 = 2.473 \cdot 10^{-5}$ is the present radiation energy density~\cite{Planck2018};
$h_{\SM,0} = 3.93$ and $g_{\gamma,0} = 2$ are the entropy and energy degrees of freedom of the SM bath
today~\cite{Saikawa:2018rcs}. The total degrees of freedom at percolation $g_{\tot,\perc}$ and $h_{\tot,\perc}$ are fixed through
$ g_{\tot,\perc} = g_{\SM,\perc} + g_{\DS,\perc}$ and $h_{\tot,\perc} = h_{\SM,\perc} + h_{\DS,\perc}$, respectively.
The normalization of the signal in eq.~\eqref{spectrum} is given by
$\tilde{\Omega} = 3 \times 0.012 \times 0.687 \times (8\pi)^{1/3} = 0.07$, with the first two factors
coming as normalization constants from ref.~\cite{Hindmarsh:2017gnf}, the third one being the overall normalization of the spectrum
$S(f)$ to unity~\cite{Caprini2019}
and the fourth one arising due to the conversion from mean bubble separations to
$\beta/H$. The efficiency $\kappa_\sw$ of converting sound waves into GWs is
calculated from $\Delta\theta/(4 \,\rho_{\DS,\perc})$, cf.~eq.~\eqref{theta}, using
the high-\vw approximation from ref.~\cite{Espinosa:2010hh}. The factor
$\mathcal{Y}$ takes into account the lifetime of the sources of GWs and
is given by~\cite{Ellis:2020awk, Caprini2019}
\begin{align}
	\mathcal{Y} = \min \bb{1, \tau_\text{sh} H} \simeq \min \bb{1, \frac{3.38}{\beta/H} \sqrt{\frac{1 + \alpha}{\kappa_\sw \, \alpha}}}.
\end{align}
The spectral shape of the GWs  is given by~\cite{Caprini2015}
\begin{align}
	S(f) = \left(\frac{f}{f_\peak}\right)^3 \ba{\frac{7}{4 + 3 \, (f/f_\peak)^2}}^{7/2},
\end{align}
with peak frequency~\cite{Bringmann:2023opz}
\begin{equation}
 f_\peak = 8.9 \, \text{mHz} \left(\frac{T_\perc}{100\text{ GeV}}\right)\left(\frac{\beta/H}{1000}\right) \ba{\frac{g_{\tot,\perc}}{100}}^{1/2}  \ba{\frac{100}{h_{\tot,\perc}}}^{1/3} \, ,
 \label{peakf}
\end{equation}
where we used that the peak frequency of the sound wave signal lies at $0.53 \,  \beta$ at the time of its emission~\cite{Hindmarsh:2017gnf}.

We will refer to a GW signal $h^2 \Omega_{\GW}(f)$ as being observable by LISA, the Einstein
Telescope~\cite{Punturo:2010zz,Maggiore:2019uih} or pulsar timing arrays~\cite{Antoniadis:2022pcn}, if it achieves a sufficiently
large signal-to-noise ratio. This means that the signal will be observable if its peak amplitude
$h^2 \Omega_{\GW}^\peak \equiv h^2 \Omega_{\GW}(f_\peak)$ lies above the power-law-integrated (PLI) sensitivity curve at
the peak frequency. The threshold signal-to-noise ratios and the PLI curves used in this work were derived in
ref.~\cite{Breitbach2019}.

\subsection{Parameter ranges}
\label{sec:scan}

The computation of properties and GW spectrum of the PT in section \ref{sec:PT} and
 \ref{sec:GWspec}, respectively, depends only on the bare parameters of the Lagrangian in
eq.~\eqref{eq:Lagrangian2}, i.e.\ the gauge coupling $g$, the quartic coupling of the dark
Higgs $\lambda$, the Yukawa coupling of the dark fermion $y$ and the vacuum expectation value of
the dark Higgs $v_\phi$. In our study we are specifically interested in exploring possible
correlations between the DM relic density and a strong GW signal in our model.

A strong GW signal typically requires sizeable $\alpha$ and values of $\beta/H$ that are not too
large. From the discussion about figure~\ref{fig:GWparams-alpha-beta}, a strong
first-order PT implies large but perturbative values of both $g$ and $\lambda$. Too small values of
these two couplings would imply very large values of $\beta/H$ and correspondingly weak GW
signals, and even cause issues of vacuum stability (for large $y$, cf.~right panel of
figure~\ref{fig:GWparams-alpha-beta}). This in turn induces an upper limit on the value of
$y$ as large values would cause an unstable vacuum for any perturbative value of $g$ and
$\lambda$. As will be seen below in section \ref{sec:relic}, successfully producing the right DM
relic density requires $m_\chi \gtrsim m_\phi$, which implies a lower limit
$y > 2\sqrt{\lambda}$. Lastly, the vev $v_\phi$ is chosen in a range that produces GWs in the
frequency range of near-future GW observatories, such as LISA.

Consequently, we randomly draw parameters from distributions that are logarithmically flat
within the following ranges: $0.1 \leq g \leq 1$, $10^{-4} \leq \lambda \leq 10^{-2}$,
$0.01 \leq y \leq 0.7$ and $10^{-3} \text{ GeV} \leq v_{\phi} \leq 10^{3}$\,GeV. We then discard
parameters that cause  the vacuum to be unstable, that do not predict a first-order PT,
or for which the PT is too supercooled and never percolates, thereby removing $82\%$ of the points drawn. The remaining
18\% of parameter points all feature a first-order PT with a corresponding GW signal. However, since the percolation
temperature is very sensitive to small changes in the couplings, the PT is only strong enough to give an observable GW signal in
certain small regions of parameter space. Indeed, only about 1\% of paramter points from the original sample feature strong supercooling ($T_p / T_c < 0.5$).

We can quantify the fine-tuning required to obtain an observable GW signal by interpreting our parameter scan as a sample drawn from the prior distributions of the parameters. We then find that out of the parameter points that give a first order PT, only about 0.8\% would be observable with LISA, whereas this number increases to 10\% if we select parameter points that give a strongly supercooled PT. For the parameter ranges that we consider (in particular of $v_\phi$) none would be observable with pulsar timing arrays or the Einstein Telescope. We note that these numbers do not correspond to rigorously calculated posterior probabilities, but rather rough estimates based on sampling densities. More precise estimates would require a different sampling strategy (see e.g.~\cite{AbdusSalam:2020rdj}), which is beyond the scope of this work.

We emphasize that this number is largely independent of the choice of priors as long as we select only parameter points that
predict any kind of first-order PT. The probability to find parameter points that give a first-order PT does however depend
sensitively on the choice of priors. If we were to extend the prior ranges for all parameters to lower couplings, the volume of
parameter space without first-order PT would grow significantly. Choosing for example $g > 0.01$ (instead of 0.1),
$\lambda > 10^{-5}$ (instead of $10^{-4}$) and $y > 10^{-3}$ (instead of 0.01) would decrease the fraction of parameter points
with a first-order PT from 18\% to 6\%. Out of these, 0.7\% would be observable by LISA, which increases to 8.5\% when considering only points with a strongly supercooled PT. As expected, our results are not very sensitive to different prior choices as we find that points that already have a first-order PT have a roughly equivalent probability of being visible at LISA regardless of the parameter ranges. In later sections, we will discuss how
these numbers change when imposing additional constraints on the dark sector, such as the relic density requirement.

Finally, when studying the
effects of thermalisation in our model in section \ref{sec:thermalization} it will be
convenient to identify a benchmark scenario with the right properties for the PT and DM
relic abundance. For reference the benchmark point is given in table \ref{tab:BP}.

\begin{table}[t]
 \centering
 \begin{tabular}{lC{0.05\textwidth}C{0.07\textwidth}C{0.07\textwidth}C{0.1\textwidth}C{0.1\textwidth}C{0.1\textwidth}C{0.1\textwidth}}
 \toprule
  & $g$ & $\lambda$ & $y$ & $v_\phi$ & $m_{\chi}$ & $m_\phi$ & $m_{A'}$\\
  \midrule
  Benchmark point & $0.67$ & $0.0035$ & $0.62$ & $430\,\text{GeV}$ & $189\,\text{GeV}$ & $36\,\text{GeV}$ & $288\, \text{GeV}$ \\
 \bottomrule
 \end{tabular}
 \caption{Benchmark point used for discussing the thermalisation of visible and dark sector.}
 \label{tab:BP}
\end{table}

\section{Dark sector relic density}
\label{sec:relic}

\begin{figure}[t!]
  \centering
  \includegraphics[width=0.95\textwidth]{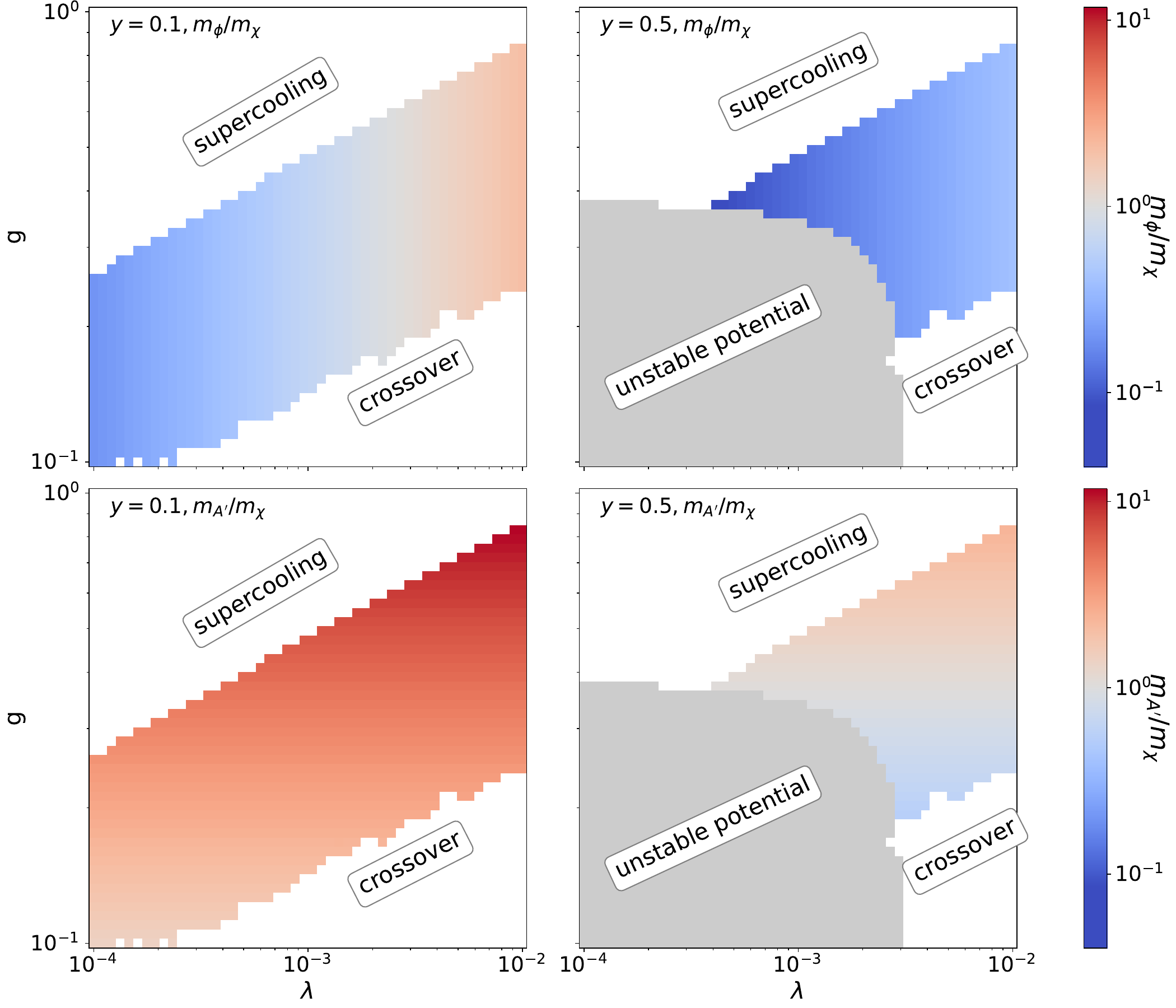}
  \caption{
  The upper (lower) panels show the ratio of the dark Higgs boson mass $m_\phi$
      (the dark photon mass $m_{A'}$) to the mass of the DM fermion $m_\chi$ for $y = 0.1$ (left) and
      $y = 0.5$ (right), as a functions of the gauge coupling $g$ and the self-interaction $\lambda$.
      Note that these ratios are independent of the dark Higgs vev.}
  \label{fig:mass_ratios}
\end{figure}

During the PT, the dark sector particles $\chi$, $\phi$ and $A'$ all obtain masses proportional
to the dark Higgs vev $v_{\phi}$. In the parameter regions of interest for a strong
first-order PT, we generally find $g > \sqrt{2\lambda}$ and $g > y/\sqrt{2}$ and hence the dark
photon is usually the heaviest state in the dark sector, cf.~eq.~\eqref{masses}. Depending
on the value of the Yukawa coupling $y$, the lightest dark sector particle will instead be
either the DM fermion or the dark Higgs boson, as shown in figure~\ref{fig:mass_ratios}.
The dark sector equilibrises soon after the PT (see section~\ref{sec:equilibrium} for a
more detailed discussion). Typically, the heaviest particles will then first drop out of
equilibrium as their number densities become strongly suppressed. The relic abundance of
the dark fermions $\chi$ is thus determined through a freeze-out process~\cite{Lee:1977ua} in
the usual way. We assume that the dark photon is unstable, decaying for example through
kinetic mixing, and therefore does not contribute to the DM relic density (unlike the case
studied in ref.~\cite{Kanemura:2023jiw}).

In our model there are three possible DM annihilation processes that are relevant for
setting the DM abundance: $\chi \chi \to \phi \phi$, $\chi \chi \to \phi A'$ and
$\chi \chi \to A' A'$. If the DM fermion is the lightest particle in the dark sector, annihilation
into other dark sector states is kinematically forbidden for vanishing kinetic energy,
such that the annihilation cross section becomes exponentially suppressed at low
temperatures. In this so-called `forbidden' regime~\cite{DAgnolo:2015ujb}, a relic
abundance in accordance with observations requires that all mass scales must be
correspondingly smaller, or the relevant couplings (much) larger. For the parameter values
we are interested in here, it is therefore typically necessary for the DM particle to be
heavier than the dark Higgs boson, which in turn requires a sizeable Yukawa coupling $y$.
For even heavier DM, with $2 m_\chi \gtrsim m_\phi + m_{A'}$, the annihilation channel
$\chi \chi \to \phi A'$ opens up. This process is a highly relevant contribution, once kinematically
accessible, as it proceeds via an $s$-wave; the annihilation into a pair of dark Higgs
bosons, $\chi\chi\to\phi\phi$, on the other hand, only proceeds via a $p$-wave.

To compute the DM relic density, we have calculated the amplitudes for all three processes, see
appendix~\ref{app:annihilation}, and implemented them in \ds~\cite{bringmann2018}, which calculates the
thermal averages and solves the full Boltzmann equation~\cite{Gondolo:1990dk}.
 While \ds\ allows precision calculations
of the relic density in a fully secluded dark sector with a varying  temperature ratio $\xi$
between the dark and the SM sector, cf.~ref.~\cite{Bringmann:2020mgx}, we will set  $\xi = 1$
for the purpose of this section. We will revisit this assumption of thermal equilibrium between the two sectors
in section~\ref{sec:equilibrium}.

\begin{figure}[t!]
  \centering
  \includegraphics[width=0.95\textwidth]{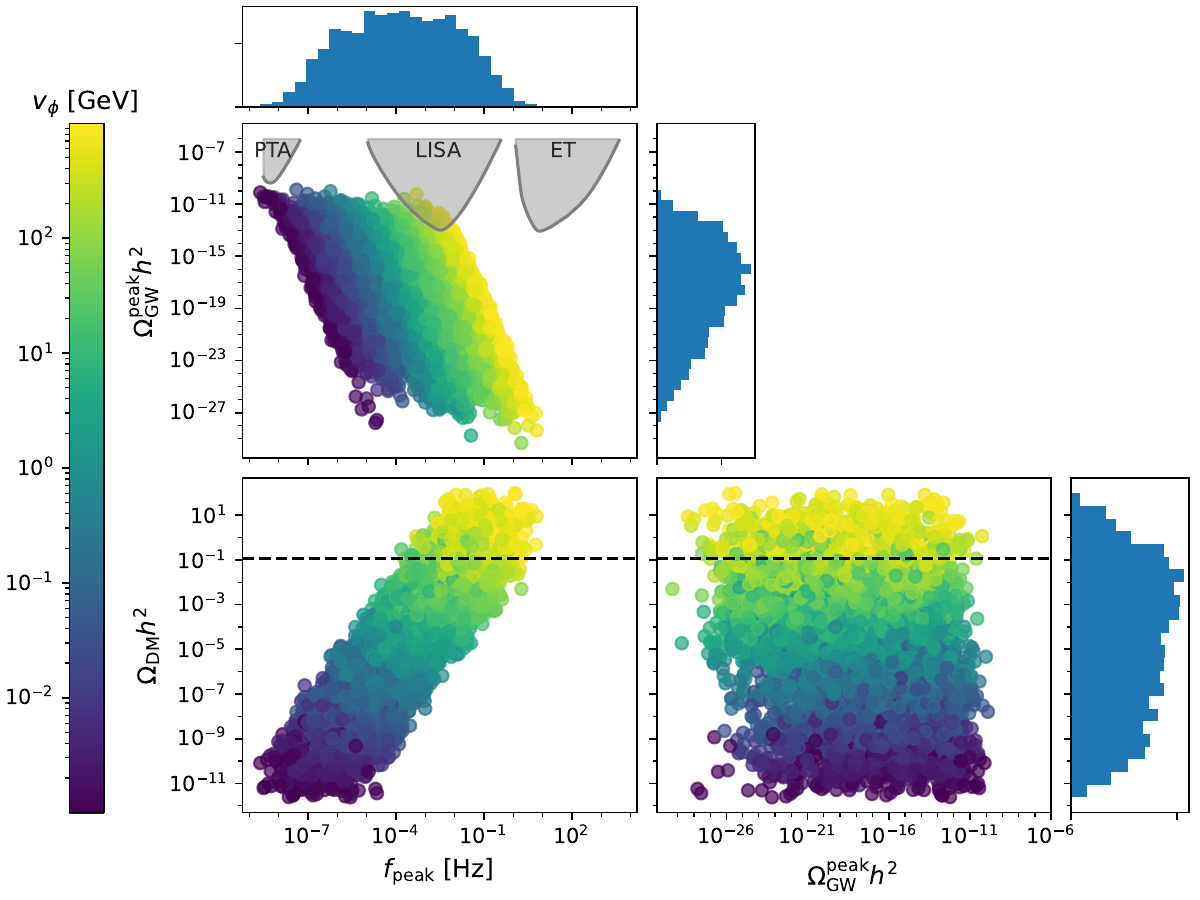}
  \caption{Scatter plots and 1D distributions of the DM density $\Omega_{\rm DM} h^2$, the GW
    density $\Omega_{\rm GW} h^2$ and the GW peak frequency $f_{\rm peak}$. For comparison, the dashed line shows the observed DM
    density, $\Omega_\DM h^2 = 0.12$~\cite{Planck2018}; grey shaded areas show the PLI sensitivities~\cite{Breitbach2019} of pulsar timing arrays, LISA and the Einstein Telescope, respectively.
  }
  \label{fig:triangle-unrestr}
\end{figure}

We show the results from the parameter scan described in section~\ref{sec:scan} in
figure~\ref{fig:triangle-unrestr}. The three two-dimensional scatter plots show the
correlation between the DM relic density $\Omega_\DM h^2$, the peak frequency $f_\peak$ well as
the peak amplitude $\Omega_\GW^\peak h^2$. One can immediately see that $\Omega_\DM$ and
$\Omega_\GW^\peak$ are not tightly correlated (with a correlation coefficient of 0.20), while
there exists a clear connection between the DM relic density and the peak frequency (with
a correlation coefficient of 0.85). We can trace this correlation back to the fact that
both quantities are determined by the dark Higgs vev $v_{\phi}$ (indicated by the colour of
each point). A smaller value of $v_{\phi}$ implies a smaller DM mass and therefore a larger
annihilation cross section, which in turn results in a smaller relic density. At the same
time, a smaller $v_{\phi}$ also implies a smaller percolation temperature, and hence a
smaller peak frequency. The strength of the PT, on the other hand, depends on the details
of the effective potential, and can vary over many orders of magnitude for any given value
of $v_{\phi}$.
We complement these scatter plots by showing distributions of the derived quantities, in the form of histograms based on
our random scan described above. For example, one can infer that most samples drawn in our setup correspond to
a peak GW signal strength of $\Omega_\GW^\peak h^2 \approx 10^{-16}$, i.e.~a few
orders of magnitude below the sensitivity of near-future GW observatories (indicated as
grey shaded areas).
Note also that the DM density caps at $\Omega_\DM h^2\approx10$, which would already correspond to an overclosed
universe; even higher values are avoided by our prior choice, in particular the upper bound on $v_\phi$ and
the lower bound on $y$.

\begin{figure}[t!]
  \centering
  \includegraphics[width=0.95\textwidth]{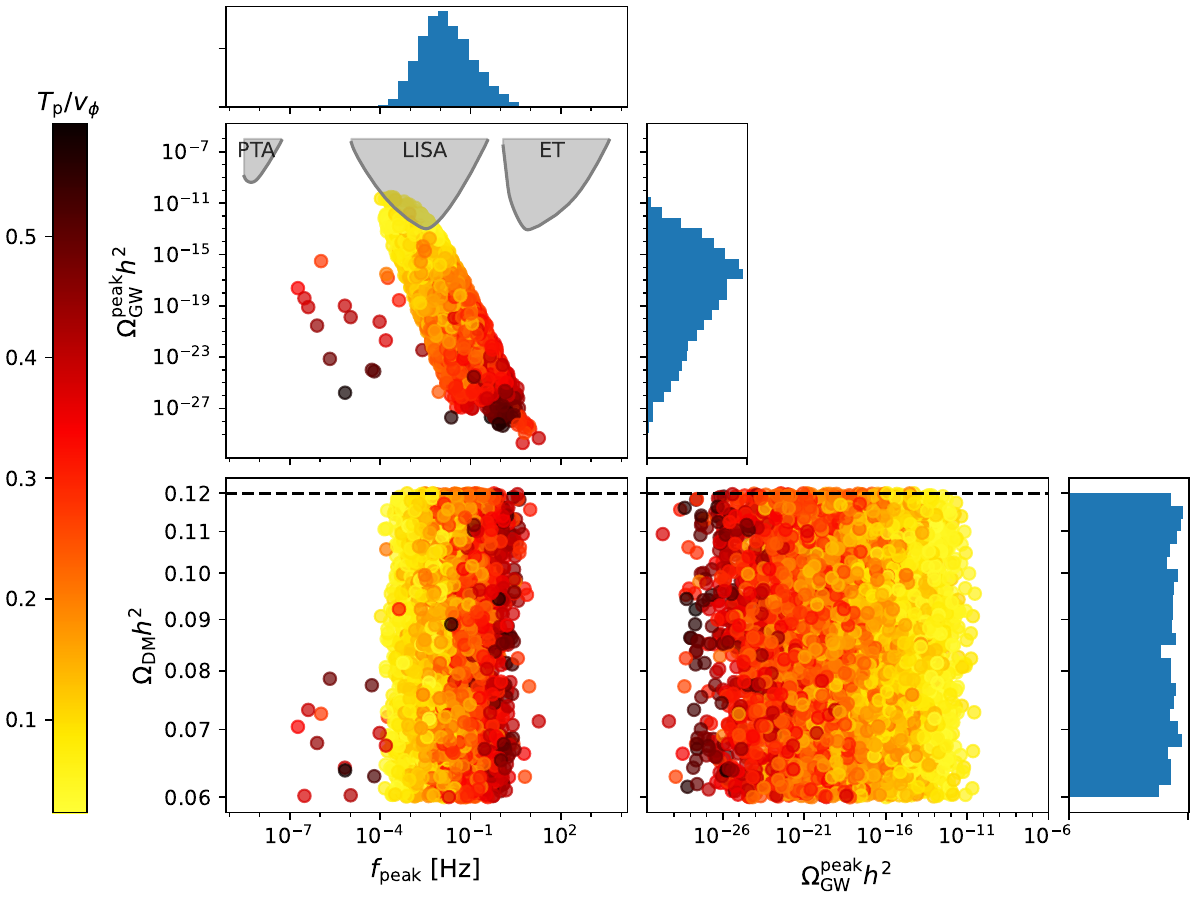}
  \caption{Scatter plots and 1D distributions as in figure~\ref{fig:triangle-unrestr}, but
    with the additional constraint $0.06 \, \leq \Omega_\DM h^2 \leq 0.12$. Here, the color
    scale does not encode the vev, but the ratio of percolation temperature
    $T_\perc$ to the vev $v_\phi$, thus indicating the amount of supercooling.
      }
  \label{fig:triangle-restr}
\end{figure}

In figure~\ref{fig:triangle-restr} we show the result of sharpening the relic density
requirement by requiring that $0.06 \, \leq \Omega_\DM h^2 \leq 0.12$. Demanding in this way that the
fermionic DM candidate in our model constitutes the dominant form of DM, the predicted
range of peak frequencies of the GW signal shrinks significantly -- as expected from the
discussion above. Interestingly, almost all viable parameter points now predict a peak
frequency between $0.1 \, \mathrm{mHz}$ and $100 \, \mathrm{mHz}$, largely overlapping
with the frequency range to which LISA is sensitive. In fact, the peak frequencies for
those parameter points that result in the strongest signal are the same as those where
LISA is most sensitive. This striking correlation is a non-trivial feature of our model
and constitutes one of our main results. Let us note that a few points remain that predict
peak frequencies outside the LISA band. Much smaller values of $f_{\rm peak}$, in
particular, correspond to parameter points in the `forbidden' regime, $m_\chi < m_\phi$, where
DM annihilations are exponentially suppressed at small temperatures. Smaller values of
$v_\phi$ (and hence smaller temperatures of the PT) can then still result in the correct DM
relic abundance, but only at the cost of significant tuning between the various couplings
(as reflected by the rareness of such parameter points, cf.~the $f_{\rm peak}$ histogram
at the top of the plot).

In contrast to figure~\ref{fig:triangle-unrestr}, where the colour coding of each point
represents the dark Higgs vev, the points in figure~\ref{fig:triangle-restr} are coloured
according to the percolation temperature of the PT, normalized to the vev $v_\phi$. Doing so
allows us to confirm that the peak amplitude of the GW spectrum is determined primarily by
the amount of supercooling. In other words, if the PT is delayed by a large potential
barrier, the strength of the PT increases, yielding strong GW amplitudes (as expected from
figures~\ref{fig:GWparams-T} and \ref{fig:GWparams-alpha-beta}). As discussed in
section~\ref{sec:scan}, the predictions for the PT properties vary a lot with small
changes of the model parameters, and thus only certain regions of the parameter space
predict a strong first-order PT. For this reason, our model cannot in general guarantee a
strong PT, and thus a GW signal that is visible with next-generation GW observatories.

We can make this statement more precise if we interpret the sampling distributions
of the model parameters as prior probabilities (as we did in section~\ref{sec:scan}),
such that the density of points in figures~\ref{fig:triangle-unrestr} and
\ref{fig:triangle-restr} can be interpreted as probability distributions for the
observables under consideration. As before, this makes it possible to quantify the
amount of fine-tuning required to obtain a strong first-order PT, through the fraction of points with a first order PT that predict a
signal observable with LISA. If we do not impose
the relic density requirement (figure~\ref{fig:triangle-unrestr}), only 0.8\% of points
with a first-order PT predict a GW signal visible at LISA, whereas this fraction increases to
3\% once the relic density requirement is included (figure~\ref{fig:triangle-restr}). If we restrict ourselves to parameter points with a strongly supercooled PT, the fraction of observable parameter points increases from 10\% to 35\%. Again, we have checked that this number is not
very sensitive to our choice of parameter ranges.

\section{Thermalisation of the two sectors}
\label{sec:equilibrium}

In this section we revisit the assumption that the temperature ratio of the dark and
visible sectors is $\xi = 1$ throughout the PT. To do so, we first need to understand the
evolution of the dark sector temperature during the PT, and convince ourselves that the
dark sector quickly thermalizes with itself afterwards, such that the dark sector states
remain in kinetic equilibrium with each other until after dark sector freeze-out (i.e.\
chemical decoupling). However, it is not necessarily the case that also the SM states are
in kinetic equilibrium with the dark sector, such that their temperature may differ from
the one of the dark sector both before and after the PT. We therefore discuss the various
processes that allow for the exchange of energy and entropy between the dark and the SM
sector, and the resulting Boltzmann equations. This enables us to identify the necessary
portal couplings for efficient thermalisation. For the case of delayed thermalisation,
after the end of the PT, we calculate the resulting dilution of the GW signal due to the
injection of entropy into the SM thermal bath.

For the purpose of illustration, we will in this section consider a specific benchmark point
that we selected from the random parameter scan discussed previously (see
table~\ref{tab:BP}). 
For $\xi=1$, the parameters of this point lead to $\alpha = 0.258$, $\beta/H=874$,
$T_\nuc=39.7\,$GeV, $T_\perc=39.1\,$GeV, $f_\peak=3\,\text{mHz}$,
$\Omega_\DM h^{2} = 0.117$, and $\Omega_\GW^\peak h^2=3\cdot10^{-13}$.
The rationale behind choosing this benchmark point is that {\it i)} the observed DM relic abundance is
reproduced (for $\xi = 1$), and that {\it ii)} the PT is sufficiently strong in order to
obtain an observable signal in LISA. We have explicitly checked that our choice is
representative in the sense that other points fulfilling these two criteria lead to a
very similar temperature evolution and resulting predictions.

\subsection{The dark sector temperature}

As the bubbles of the broken phase expand, more and more dark sector particles will pass
through the bubble walls and enter the new phase. In the process, not only their rest masses
but also their kinetic energies increase dramatically, by converting the vacuum energy of the dark
Higgs field stored in the false vacuum.
Here we neglect the small fraction of the energy density that is converted into GWs and assume
that the bubble walls have already reached their terminal velocity, such that no energy is
needed for their acceleration. As we have learned in the previous section, in particular, the energy
density of  GWs produced in the PT is bounded by
$\Omega_\GW^\peak h^2 < 10^{-10}$ and can therefore safely be ignored.
We also neglect the effect of bubble filtering~\cite{Baker2019, Chway2019}, i.e.\ we assume that all dark sector particles can enter the new phase. This is a good approximation for sufficiently fast bubble walls, see appendix~\ref{app:darkwalls} for details.

Since the different particle species in the dark sector were all relativistic before the
PT, their number densities immediately after
the PT will  be comparable, even though their masses will now be very different. Indeed, for
strongly supercooled PTs the dark photons (and possibly also the DM particles) will typically
have a large mass  compared to the temperature of the
plasma, such that their equilibrium number density would be Boltzmann-suppressed. In
other words, right after the PT the dark sector finds itself far away from thermal
equilibrium. Nevertheless, interactions between the different dark sector particles are
rather strong, and hence the heavier particles are expected to annihilate rapidly into
lighter ones, thereby restoring equilibrium.

As we will show below, the time required to reach equilibrium is negligible compared to
the duration of the PT, such that we can to a very good approximation define a dark sector
temperature of the broken phase $\TDSbro$ immediately after the PT.
This temperature is obtained from the temperature of the symmetric dark sector phase
$\TDSsym$ using energy conservation:
\begin{equation}
  \rho_\text{vac}(\TDSbro) + \rho_\DS(\TDSbro) =  \rho_\text{vac}(\TDSsym) + \rho_\DS(\TDSsym) \, , \label{eq:energy_cons}
\end{equation}
where $\TDSbro$ denotes the temperature in the broken phase and
\begin{equation}
  \rho_\DS(\TDSbro) = \frac{\pi^2}{30} \, g_{\DS}(\TDSbro) \, \left(\TDSbro\right)^4 ,
\end{equation}
where $g_{\DS}(T)$ takes into account the $T$-dependence stemming both from
thermal (field-dependent) masses and the minimum of the effective potential.
Eq.~\eqref{eq:energy_cons} can easily be solved numerically for $\TDSbro$ for a given $\TDSsym$ .

In practice, we find that slightly different temperatures $\TDSbro$ of the broken phase
are obtained when solving the equation taking $\TDSsym= T_\perc$ or $\TDSsym =
T_\nuc$. This is because
the energy density of the broken phase redshifts differently from the symmetric phase. We
have therefore implemented a more detailed calculation, which tracks the temperature of
the symmetric and broken phases from bubble nucleation to percolation and applies
eq.~\eqref{eq:energy_cons} at each time step to the fraction of the universe entering the
broken phase. Here the energy in the bubble walls, which for relativistic bubble wall velocities
redshifts like radiation~\cite{Gouttenoire:2023naa}, is included in the energy in the
symmetric phase. We find that this more careful treatment gives very similar results to simply
applying eq.~\eqref{eq:energy_cons} at $\TDSsym = T_\perc$. We therefore use the latter prescription in the following when computing the temperature of the dark sector after the phase transition.

\begin{figure}[t]
  \centering
  \includegraphics[width=0.65\textwidth]{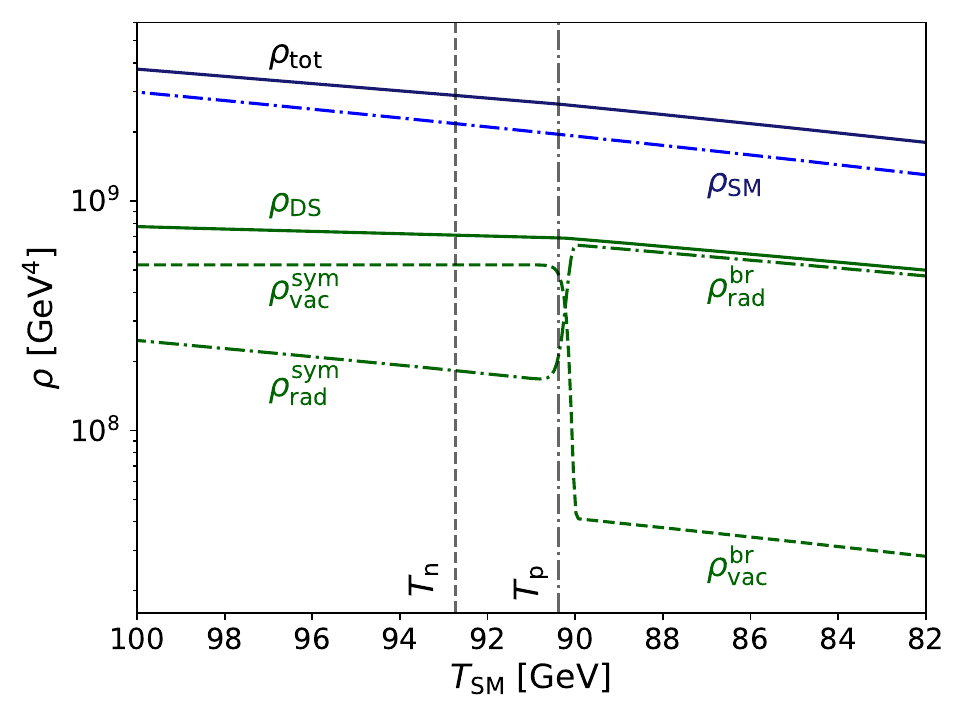}
  \caption{Contributions to the energy density around the PT for our
    benchmark point, as a function of the SM temperature and for a temperature ratio $\xi = 1$.
     The energy densities in the symmetric and broken phase of the dark sector have
    two contributions, namely the energy of the particles (`rad') and the potential energy of the
    scalar field (`vac').
    }
  \label{fig:hubble-parameter}
\end{figure}

For our benchmark point, we find that the energy density of the dark sector before the PT
is dominated by vacuum energy
(see figure~\ref{fig:hubble-parameter}). In the broken phase, on the other hand, the
vacuum energy is very small and quickly relaxes to a value very close to its zero temperature value
as the
temperature decreases further.
This difference in vacuum energy leads to a substantial
reheating of the dark sector, which, as a result, is 
hotter than the SM sector after the PT. For our benchmark point, we find that if the two sectors have equal
temperature before the PT, in the broken phase the dark sector temperature will be larger by a factor of about 1.3.
This reheating of the dark sector typically ensures that the
dark Higgs bosons will be relativistic immediately after the PT.

\subsection{Thermalisation within the dark sector}
\label{sec:thermalization}

In the discussion above we have assumed that the dark sector can be characterised by a
common temperature shortly after the PT. To justify this approach, we need an estimate of
the time $\tau$ required to reach this equilibrium state and show that it is sufficiently
small. For this purpose, we calculate the interaction rate for each dark sector state $X$
in thermal equilibrium:
\begin{equation}
  \Gamma_X = \sum_Y \langle \sigma_{XY} v \rangle n_Y^{\rm eq} \,,
\end{equation}
where the sum is over all dark sector states $Y$, $\sigma_{XY}$ denotes the total interaction
cross section of $X$ and $Y$, brackets denote thermal averaging
(for simplicity calculated by assuming Boltzmann distributions)
and $n_Y^{\rm eq}$ denotes the equilibrium number
density of $Y$. 

A total of 20 different processes contribute to the thermalisation of the dark sector, the
relative importance of which depends on the specific choice of parameters and the dark
sector temperature. In the interest of brevity we refrain from stating the thermalisation
rates explicitly. Broadly speaking, we find that $\Gamma_X$ is only a few orders of magnitude
smaller than $m_X$. For example, dark Higgs bosons can thermalise via self-scattering,
i.e.\ $\phi \phi \leftrightarrow \phi \phi$, for which the scattering cross section is
$9 \lambda^2 / (8 \pi s)$. For temperatures comparable to the dark Higgs boson mass, we have
$s \approx 4 m_\phi^2$ and $n_\phi \approx \zeta(3) m_\phi^3 / \pi^2$, such that
$\Gamma_\phi \sim 10^{-7} m_\phi$ for the benchmark point. Interactions of the dark Higgs bosons with
dark fermions or dark photons benefit from the larger couplings $y, g \gg \lambda$, but suffer
from a Boltzmann suppression if $T_\perc < m_\chi, m_{A'}$.

A rough estimate of the thermalisation timescale is then obtained via 
\begin{equation}
  \tau = \max_X \, \Gamma_X^{-1} \, .
\end{equation}

\begin{figure}
  \centering
  \includegraphics[width=0.95\textwidth]{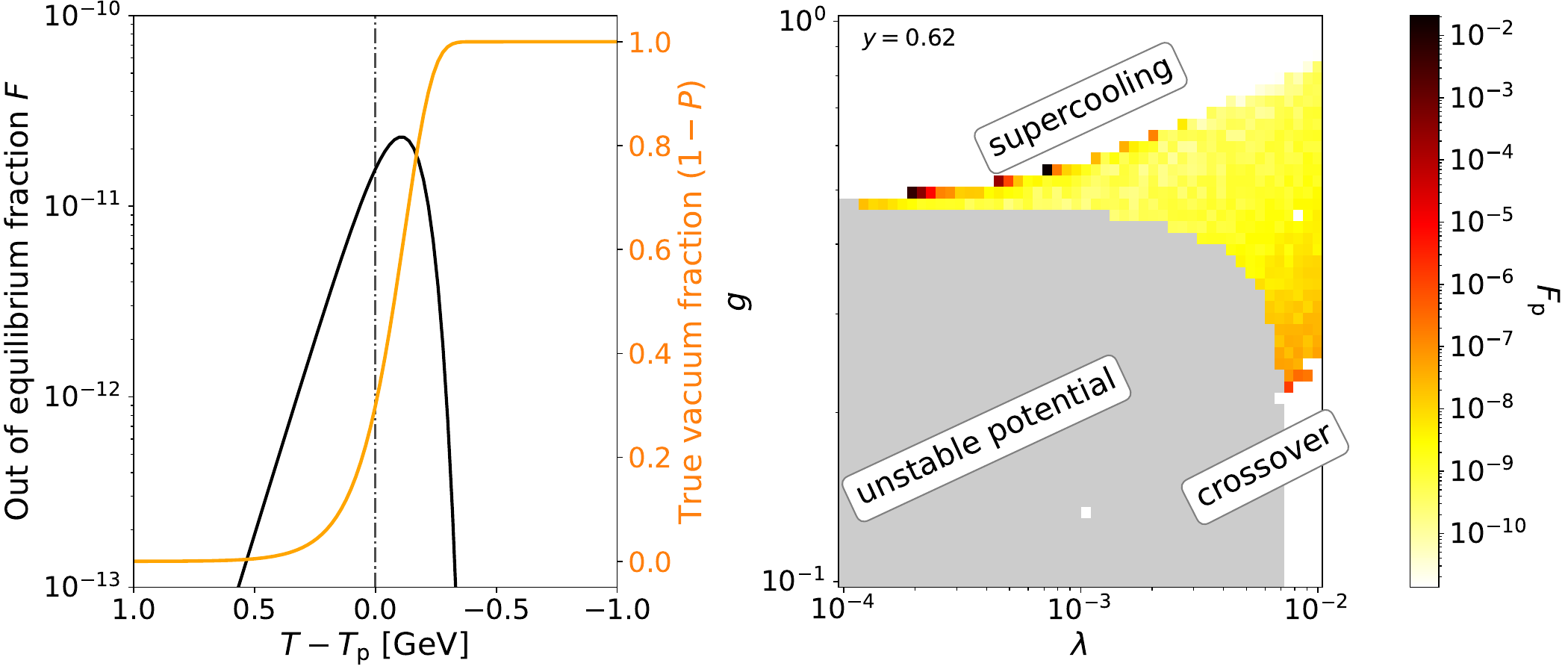}
  \caption{{\it Left:} The black line shows the energy fraction of the dark sector that is out of
    equilibrium for our benchmark point, as a function of the SM photon temperature.
    For reference, the fraction of the dark sector in the broken phase is given by the
    orange line and the percolation temperature as the dashed vertical line. {\it Right:}
    Out-of-equilibrium fraction as a function of $\lambda$ and $g$ for $y = 0.62$ and
    $v_\phi=430$\,GeV.}
  \label{fig:thermalisation}
\end{figure}

Given the timescale $\tau$ we can estimate the out-of-equilibrium fraction of the universe
$F(t)$, which describes the relative volume of the universe that entered the broken phase
so recently that it has not had enough time to reach thermal equilibrium. The related
false vacuum fraction $P(t)$, cf.~eq.~\eqref{eq:falsvacfrac}, describes the fraction of
the universe which has not yet transitioned to the new phase, such that the true vacuum
fraction is given by $P_\true(t) = 1 - P(t)$. Its time derivative $\dot{P}_\true$
describes the rate with which the volume is transitioning to the true minimum of the
potential for a given time $t$. We introduce the quantity
\begin{equation}
  F(t) \equiv P(t - \tau) - P(t) > 0\,,
\end{equation}
which can hence be interpreted as the volume fraction $\dot{P}_\true \, \Delta t$ that just
transitioned to the broken phase within the small thermalization time scale
$\Delta t = \tau$, cf.~figure~\ref{fig:hubble-parameter}. The volume fraction $F$ becomes small
exactly when the thermalisation timescale $\tau$ is small compared to the transition
timescale $1/\beta$, as can be seen from the saddle point approximation of $P(t)$ around the
percolation time $t_\perc$:
\begin{align}
  \begin{split}
  F(t) &\approx \exp \left( -0.34 e^{\beta(t - t_\perc - \tau)} \right) -
  \exp \left( -0.34 e^{\beta(t-t_\perc)} \right)  \\
    &\approx \beta \tau e^{\beta(t - t_\perc)} \exp \left( -0.34 e^{\beta(t - t_\perc)} \right) \leq 0.37 \beta\tau\, .
  \end{split}
\end{align}
Here, the last term follows by inserting the time at which $F(t)$ peaks, which is found to
be $t \approx t_{\perc} - 1.08/\beta$. Alternatively, one can interpret $F$ as the volume fraction
of a shell around the bubbles with the width of the mean free path of the particles that
just entered the bubbles. In the left panel of figure~\ref{fig:thermalisation} we show $F$
as a function of $T - T_\perc$ for our benchmark point. As expected, we find that $F$
takes its maximal value close to the percolation temperature $T_\perc$. Nevertheless, even
at the peak the value is tiny, implying that the fraction of the universe that is not in
thermal equilibrium and therefore cannot be described by a temperature is completely
negligible. In the right panel of figure~\ref{fig:thermalisation} we show the maximum
value $F_\perc$ at the percolation temperature for more points from our parameter scan,
varying both $g$ and $\lambda$.
Similarly we find that for all our cases the thermalisation is fast compared to the
transition timescale. Even strongly supercooled points, where $F_\perc$ becomes close to
unity, are expected to return to equilibrium before freeze out, since $\beta/H \gg 1$.

So far, we have throughout assumed that chemical potentials in the dark sector can be
neglected after thermalization has taken place. This is certainly a good
approximation as long as at least the lightest state in the dark sector has a mass that is
small compared to the dark sector temperature, which is typically the case shortly after
the PT. However, as the universe continues to cool down, this assumption becomes
increasingly critical. If we assume that the dark sector cannot transfer its entropy to
the SM thermal bath, the subsequent evolution depends crucially on whether number-changing
processes of the lightest state, such as $3\phi \to 2\phi$, are efficient enough to
maintain chemical equilibrium in the dark sector. If this is the case, the dark sector
temperature will decrease much more slowly than the SM temperature, and the universe will
enter a period of `cannibal' domination~\cite{Pappadopulo:2016pkp,Farina:2016llk}.
If, on the other hand, number-changing processes
are inefficient, the dark sector will develop large chemical potentials and the universe
will eventually enter a period of matter domination. In both cases, the energy and entropy stored in
the dark sector must later be transferred to the SM heat bath in order to recover
radiation domination before neutrinos decouple at $T \approx 2 \, \text{MeV}$, marking the onset of big bang nucleosynthesis~\cite{Bringmann:2023opz}. Neither of these scenarios is very
desirable, as the GW signals from the PT will be strongly diluted in the process~\cite{Ertas:2021xeh}.

We are therefore more interested in the case where the dark and SM sector quickly
equilibrate after the PT, such that their temperatures become equal, chemical
potentials become negligible, and the universe evolves approximately as in radiation
domination. In the following we discuss the processes that contribute to this
process, and we derive the coupling strengths required for it to happen rapidly enough.

\subsection{Thermalisation of the dark and visible sector}

The conceptually simplest way for the dark and visible sector to exchange entropy and
energy is via Higgs
mixing~\cite{Schabinger:2005ei,Patt:2006fw,Weihs:2011wp,Duerr:2016tmh,Li:2023bxy}.\footnote{%
  A second possibility would be to consider kinetic mixing between the dark photon and SM
  hypercharge. However, given that the dark photon is typically the heaviest state in the
  dark sector, it will be strongly Boltzmann-suppressed at low temperatures, and therefore
  cannot efficiently keep the two sectors in equilibrium. } Such a mixing arises from an
additional term in the scalar potential:
\begin{equation}
  V_\text{mix}(H, \Phi) = \lambda_{h\phi} |H|^2 |\Phi|^2 \, .
\end{equation}
As long as the vevs of both Higgs bosons vanish, the dominant process connecting the two
sectors is $H H \to \Phi \Phi$. As soon as one of the two Higgs bosons acquires a vev, it can
decay into the other one, i.e.\ $h \to \Phi \Phi$ (if the electroweak symmetry breaks first) or
$\phi \to H H$ (if the dark symmetry breaks first). If kinematically allowed, these decay
processes typically dominate over the $2 \to 2$ process for non-relativistic particles.

After both electroweak symmetry and the dark gauge symmetry have been spontaneously
broken, the Higgs mixing generates a non-diagonal mass term, which can be rotated away by
introducing the mixing angle
\begin{equation}\label{eq:mixing-angle}
  \theta = \frac{\lambda_{h\phi} v_\phi  v_h}{m_h^2 - m_\phi^2} \, ,
\end{equation}
where we have assumed $\theta \ll 1$ both in order to satisfy experimental constraints on the properties of the SM-like Higgs boson and to ensure that thermal corrections from SM fields to the effective potential are negligible so that the dark sector PT can be treated separately from the EWPT.
Both the masses and the vevs, and hence also the mixing angle,
depend on the temperature. As a result of this mixing, the dark Higgs boson obtains
couplings to SM fermions and gauge bosons proportional to $\theta$. Of the greatest relevance
for our discussion will be the decay of dark Higgs bosons into bottom quarks $b$, with a tree-level decay
width given by
\begin{equation}
  \Gamma_{\phi \to b\bar{b}} = \frac{3  m_\phi m_b^2 \sin^2 \theta}{8\pi v_h^2} \sqrt{1 - \frac{4 m_b^2}{m_\phi^2}} \, .
\end{equation}

To calculate the entropy transfer between the dark and visible sector, we define the
heat transfer rate
\begin{equation}
  \dot{q} \equiv \dot{\rho} + 3 H (\rho + P) \,,
\end{equation}
which is related to the change of entropy density via
\begin{equation}
  \dot{s} = -3  H  s + \frac{\dot{q}}{T} \,.
\end{equation}
At the same time, the first moment of the Boltzmann equation gives
\begin{equation}
  \int \frac{\mathrm{d}^3 p}{(2\pi)^3} E\, L[f] = \dot{q} =  \int \frac{\mathrm{d}^3 p}{(2\pi)^3} E\, C[f] \, .
\end{equation}
The general expression for the first moment of the collision operator for decays
(including relativistic corrections and quantum effects)
was derived in refs.~\cite{Bringmann:2021sth, Bernreuther:2022bdw}.
It was shown there that the leading relativistic effects (namely a time dilation of the decay
proportional to $1/\gamma$ and an increase of the injected energy by a factor $\gamma$) cancel,
and it is therefore a good approximation to assume that the decaying particle is at rest.
The integral of the collision operator thus gives
\begin{equation}
  \dot{q} \simeq m_\phi \left(\dot{n} + 3 H n\right) \,,
\end{equation}
and the evolution of the number density is given by the usual Boltzmann equation
\begin{equation}
  \dot{n}_\phi + 3  H  n_\phi = - \Gamma_{\phi \to b \bar{b}}  \, n^\text{eq}_\phi \left(\frac{n_\phi}{n^\text{eq}_\phi} - 1 \right) .
\end{equation}
Here $n^\text{eq}$ denotes the equilibrium number density for $T_\DS = T_\SM$,
whereas the actual number density can be calculated from the dark sector temperature and
the assumption of negligible chemical potential.
Putting everything together, we obtain
\begin{equation}
  \dot{s}_\DS = - 3 H  s_\DS - \frac{m_\phi}{T_\DS} \, \Gamma_{\phi \to b \bar{b}} \, n^\text{eq}_\phi
  \left( \frac{n_\phi}{n^\text{eq}_\phi} - 1 \right).
\end{equation}
A completely analogous equation holds for the evolution of the SM entropy density
$s_\SM$. Since the Hubble rate $H$ depends on the combined energy density of both
sectors, both equations need to be solved simultaneously, together with the equation
$\dot{a} = H a$, which we will need below to calculate the evolution of the GW spectrum.

In practice, we also include decays into lighter quarks and leptons, which become relevant
if decays into bottom quarks are kinematically forbidden. We further include the processes
$h \to \phi \phi$ and $\phi \to h h$ if they are kinematically allowed.
We do not, however, include $2 \to 2$ processes of the form $q \bar{q} \to g \phi$ or $q \phi \to q g$,
which may give a non-negligible contribution for light dark Higgs
bosons~\cite{Evans:2017kti,Fradette:2018hhl}. Additional details, and the relevant
equations, can be found in appendix~\ref{app:thermalisation}.
We note that
analogous equations to the ones above can be derived for the case that only one Higgs
boson has a vev and the case that both symmetries are unbroken.

\begin{figure}[t]
  \centering
  \includegraphics[width=0.95\textwidth]{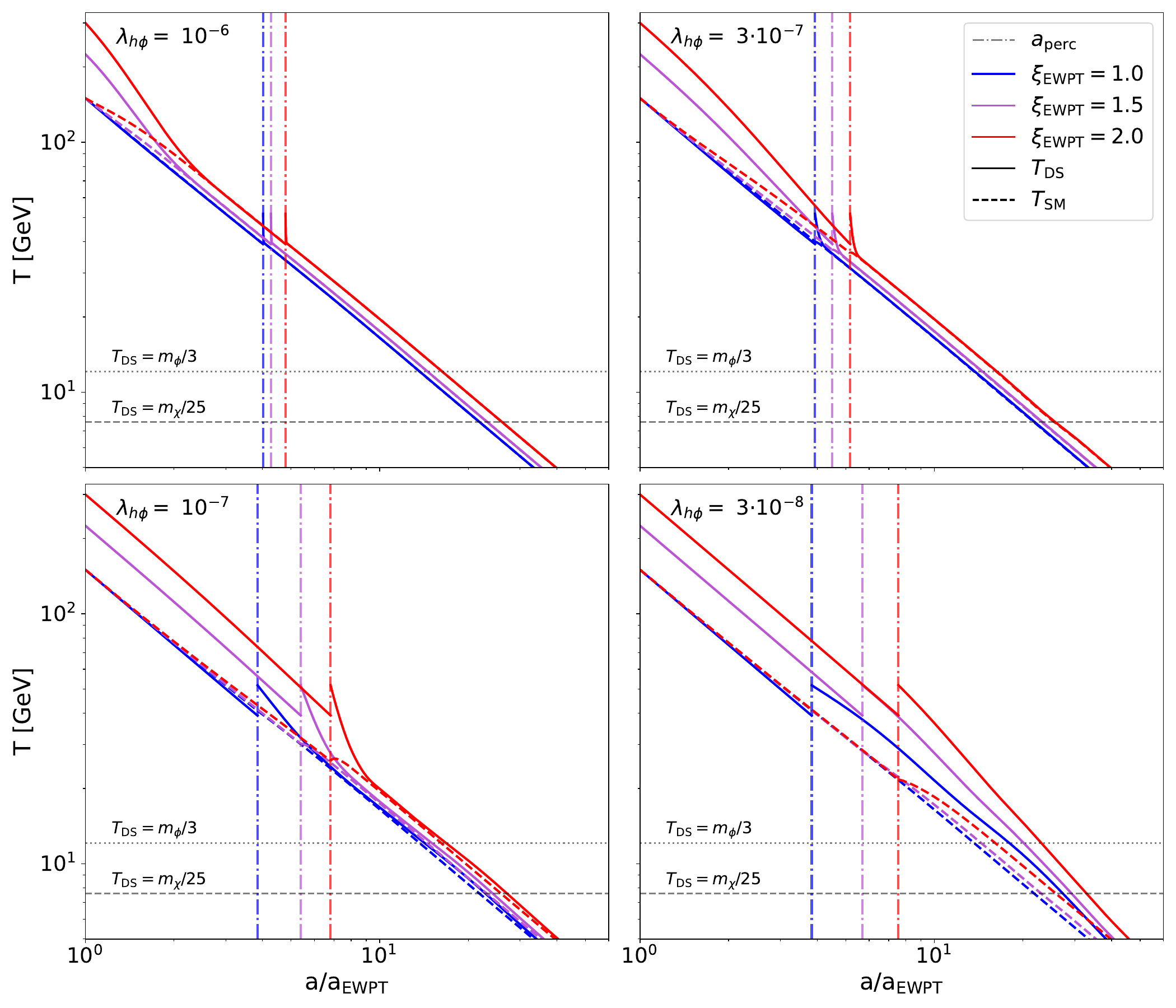}
  \caption{Evolution of the dark sector temperature (solid) and the SM temperature
    (dashed) for different values of the initial temperature ratio $\xi_\EWPT$ as a function
    of scale factor. The vertical lines indicate the scale factor at percolation, which
    depends on the temperature ratio. The horizontal lines indicate when the lightest
    state in the dark sector becomes non-relativistic ($T_\DS = m_\phi / 3$) and
    approximately when the DM particles freezes out ($T_\DS = m_\chi / 25$). The different
    panels correspond to different values of the portal coupling $\lambda_{h\phi}$. }
  \label{fig:evolution}
\end{figure}

As discussed above, it is not necessarily the case that the dark sector is in kinetic
equilibrium with the SM at high temperatures. In the following, we will therefore take the
temperature ratio of the two sectors at the electroweak phase transition (EWPT),
$\xi_\EWPT$, as a free parameter and calculate the evolution of $\xi$ during the subsequent
cosmological stages (see appendix~\ref{app:thermalisation} for details). In
figure~\ref{fig:evolution} we show the visible and dark sector temperatures as a function
of scale factor for different initial values of $\xi_\EWPT$, defined at
$T_\SM = 150 \, \text{GeV}$. In the four panels the portal coupling was set to the
representative values $\lambda_{h\phi} = 10^{-6}$ (top left), $3 \cdot 10^{-7}$ (top right),
$10^{-7}$ (bottom left) and $3 \cdot 10^{-8}$ (bottom right). In each panel, we indicate the
moment of percolation by a vertical dot-dashed line; the approximate temperature of the
dark Higgs becoming non-relativistic ($m_\phi / T_\DS = 3$) and the DM fermion freezing out
($m_\chi / T_\DS = 25$) are indicated by horizontal dotted and dashed lines, respectively. We
make the following observations:
\begin{itemize}
\item For $\lambda_{h\phi} = 10^{-6}$, the two sectors thermalise efficiently already before
  percolation. The initial value of $\xi_\EWPT$ is therefore inconsequential for the
  subsequent evolution, and we obtain the same results for all cases.
\item For $\lambda_{h\phi} = 3 \times 10^{-7}$, the two sectors do exchange energy and entropy already
  in the unbroken phase, but do not fully thermalise before percolation, such that
  $\xi_\perc$ depends on $\xi_\EWPT$. After dark symmetry breaking, the two
  sectors thermalise rapidly, so that the subsequent evolution, and in particular the
  relic density calculation, do not depend on $\xi_\EWPT$.
\item For $\lambda_{h\phi} = 10^{-7}$, the energy exchange before dark symmetry breaking is
  completely negligible. Even after dark symmetry breaking, it will take a while for the
  temperatures of the two sectors to approach each other. Nevertheless, the two sectors
  reach equilibrium before the dark Higgs bosons become non-relativistic.
\item For $\lambda_{h\phi} = 3 \times 10^{-8}$, the two sectors do not quickly thermalize after the
  PT, and the universe enters an early period of cannibal
  domination.\footnote{For the purpose of this plot, we assume that number changing
    processes in the dark sector remain efficient throughout, such that the chemical
    potential of the dark Higgs boson vanishes and (in the absence of decays) the dark
    sector temperature grows relative to the SM temperature.}
\end{itemize}
In general the value of $\lambda_{h\phi}$ needed to ensure thermalisation depends somewhat on the
dark Higgs vev, since for $m_\phi \ll m_h$ the mixing angle scales as
$\theta \propto \lambda_{h\phi} v_{\phi}$. Moreover, for small dark Higgs boson masses, decays into bottom quarks
are kinematically forbidden and thermalisation is less efficient. For the parameter points
that reproduce the observed relic density we find that the assumption $\xi = 1$ made in
section~\ref{sec:relic} is well justified for $\lambda_{h\phi}$ greater than
$10^{-6}$--$10^{-5}$. This value should be compared to the currently strongest bounds from
direct detection experiments, which are only sensitive to
$\lambda_{h\phi} \gtrsim 10^{-3}$~\cite{Ferber:2023iso}.

\section{Gravitational waves from hot dark sectors}
\label{sec:hot}

While the analysis of the dark sector PT is simplest for $\lambda_{h\phi} > 10^{-6}$, it is
phenomenologically interesting to also consider smaller values of $\lambda_{h\phi}$, such that the
temperature ratio of the two sectors before the PT may differ from unity.
The reason is that $\xi_\perc> 1$ leads to an enhancement of the GW signal, as a
result of the larger total energy density in the dark sector compared to the SM thermal
bath~\cite{Breitbach2019,Ertas:2021xeh}. In this case, however, we also need to consider what happens
to the energy density of the dark sector after the PT.

If the transfer of energy from the dark to the visible sector after the PT is slow, the
energy density of the universe will eventually be dominated by non-relativistic dark sector particles.
This effect is already visible in figure~\ref{fig:evolution}, where for small values of
$\lambda_{h\phi} = 3 \times 10^{-8}$ the temperature ratio $\xi$ still differs from unity when the
lightest dark sector particle becomes non-relativistic.
When the non-relativistic dark Higgs bosons eventually decay into SM particles, their entropy is
transferred to the thermal bath; this modifies the expansion history of the universe and leads to a
dilution of the GW signal. It is a priori unclear whether this dilution effect dominates
over the enhancement with increasing dark sector temperature, or whether a net increase in the
strength of GW signals remains. Moreover, the dilution effect also shifts the GW frequencies and
might thereby spoil the correlation between peak frequency and relic density found in
section~\ref{sec:relic}. In the following we will investigate these effects in detail.

If the portal coupling is extremely small, in principle even the relic density calculation could be
modified. If the dark Higgs bosons become non-relativistic after freeze-out, in particular,
they may come to dominate the energy density of the universe at 
later times, and dilute not only the GW energy density but
also the DM energy density through their decays (see e.g.~ref.~\cite{Berlin:2016gtr}).
If, on the other hand, the dark Higgs bosons are non-relativistic already during freeze-out, 
inefficient thermalisation between the two sectors may additionally
imply a non-trivial evolution of the dark sector temperature during
freeze-out. While these are interesting scenarios in their own right, they are beyond the scope of 
this work. Instead,  we will here  focus exclusively on the case where the dark Higgs bosons decay
sufficiently quickly for the standard freeze-out calculation
(with temperature ratio $\xi = 1$) to be valid.

\subsection{Dilution of gravitational waves}
\label{sec:dilution}

\begin{figure}[t]
  \centering
  \includegraphics[width=0.65\textwidth]{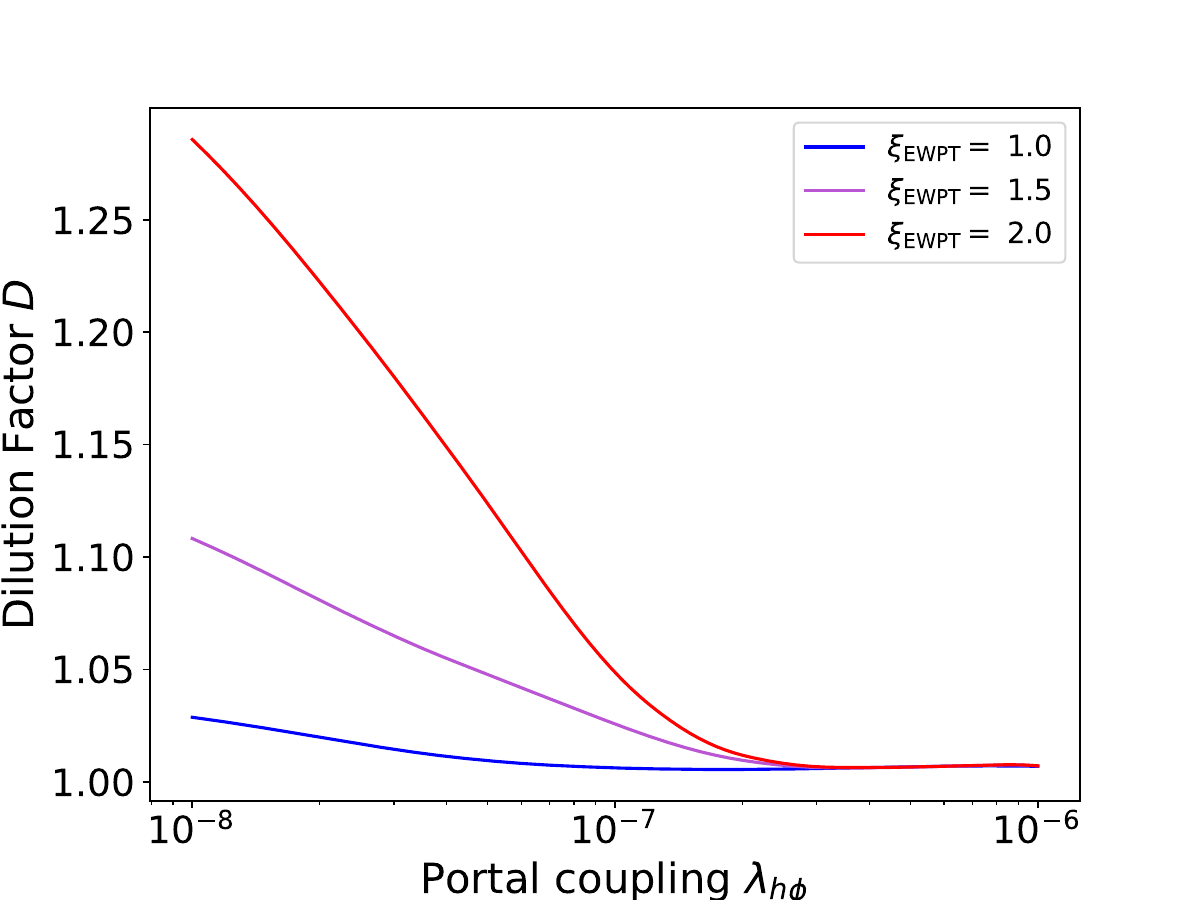}
  \caption{Dilution factor $D$, cf.~eq.~\eqref{eq:redshift},
  as a function of the portal coupling $\lambda_{h\phi}$, and for different
  values of the initial temperature ratio $\xi_\EWPT$ as indicated.
   }
  \label{fig:dilution}
\end{figure}

We described the GW spectrum in section~\ref{sec:GWspec} under the assumption of equal initial temperatures 
between the dark and SM sectors and the absence of any entropy injection into the SM bath. If the temperatures are unequal, the 
total degrees of freedom used in eqs.~\eqref{redshift} and \eqref{peakf} are modified as~\cite{Husdal:2016haj}
\begin{align}
	g_{\tot,\perc} &= g_{\SM,\perc} + g_{\DS,\perc} \, \xi_\perc^4 \,, \label{eq:gtotrho}\\
	h_{\tot,\perc} &= h_{\SM,\perc} + h_{\DS,\perc} \, \xi_\perc^3 \,,  \label{eq:gtots}
\end{align}
where the temperature ratio at percolation $\xi_\perc$ is derived from the initial temperature ratio $\xi$ by assuming entropy 
conservation in the dark sector, which is valid before the onset of thermalization of the two sectors (see section~\ref{sec:thermalization}).
Gravitational waves redshift like radiation, such that the spectrum $\Omega_\GW$
observed today can be calculated from the emitted spectrum $\Omega_\GW^\text{em}$ as
\begin{align}
	\Omega_\GW (f) & = \left(\frac{a_\perc}{a_0}\right)^4 \left(\frac{H_\perc}{H_0}\right)^2 \Omega_\GW^\text{em} \left(\frac{a_0}{a_\perc} \, f\right) \equiv \mathcal{R} \, \Omega_\GW^\text{em} \left(\frac{a_0}{a_\perc} \, f\right) . \label{eq:R}
\end{align}
Here we assume that all modes redshift equally and thus ignore possible spectral features,
which can for instance arise in the case of an early matter
domination~\cite{Barenboim:2016mjm}. This is reasonable as we are only interested in the
GW spectra close to their peak.
Assuming
entropy conservation between the end of the phase transition and today, we have
\begin{align}
	\frac{a_0}{a_\perc} = \left(\frac{h_{\SM,\perc}}{h_{\SM,0}}\right)^{1/3} \left(\frac{T_{\SM,\perc}}{T_{\SM,0}}\right) .\label{eq:scalefactorratio}
\end{align}
Together with eq.~\eqref{eq:R} this directly yields the previously used eq.~\eqref{redshift}.
If entropy is however not conserved in the SM bath at early times, the ratio of scale factors
$a_0 / a_\perc$ in eq.~\eqref{eq:scalefactorratio} needs to be corrected by a factor of
$D_\SM^{1/3}$, where $D_\SM = S_{\SM,0} / S_{\SM,\perc}
$~\cite{Cirelli2018, Ertas:2021xeh}. We hence find, more generally, that
\begin{align}
	\mathcal{R}h^2 &=  \frac{\Omega_\rad h^2}{D_\SM^{4/3}}\, \ba{\frac{h_{\SM,0}}{h_{\SM,\perc}}}^{4/3} \frac{g_{\tot,\perc} }{2}
	=  \frac{\Omega_\rad h^2}{D^{4/3}}\, \ba{\frac{h_{\SM,0}}{h_{\tot,\perc}}}^{4/3}  \frac{g_{\tot,\perc}}{2}
	\,,
	\label{eq:redshift}
\end{align}
where we introduced the more convenient dilution factor 
$D\equiv D_\SM \, h_{\SM,\perc} / h_{\tot,\perc}$~\cite{Ertas:2021xeh,Cirelli2018}. Analogously, the peak frequency obtains an 
additional redshift for the case of a non-conserved SM entropy and hence reads
\begin{align}
		f_\peak = \frac{8.9 \, \text{mHz}}{D^{1/3}} \left(\frac{T_{\SM,\perc}}{100\text{ GeV}}\right)\left(\frac{\beta/H}{1000}\right)  \ba{\frac{g_{\tot,\perc}}{100}}^{1/2}  \ba{\frac{100}{h_{\tot,\perc}}}^{1/3} \, .
\end{align}

We show in figure~\ref{fig:dilution} the dilution factor $D$, as a function of the portal
coupling $\lambda_{h\phi}$. Here, we choose the same benchmark point as studied in
section~\ref{sec:equilibrium}, and show the result for different values of the initial
dark sector temperature ratio $\xi_\EWPT$, defined at $T_\SM = 150\,\text{GeV}$. As expected, for
sufficiently large $\lambda_{h\phi}$ there is no significant dilution\footnote{%
In the limit of large portal couplings $\lambda_{h\phi}$, the dilution factor $D$ approaches a value slightly larger than 1. This is 
an expected feature, indicating a negligible dilution effect that is entirely due to the additional degrees of freedom in the 
combined thermal bath of SM and dark sector particles and not a consequence of  additional entropy injection (with respect to
$\Lambda$CDM) into the SM bath after the PT.
}, 
as entropy is conserved and
only radiation degrees of freedom contribute to the energy content of the universe.
However, with decreasing $\lambda_{h\phi}$ this is no longer the case and the dilution
factor grows, becoming sizeable for $\lambda_{h\phi} \ll 10^{-7}$.

\subsection{Results}

\begin{figure}[t]
  \centering
  \includegraphics[width=0.95\textwidth]{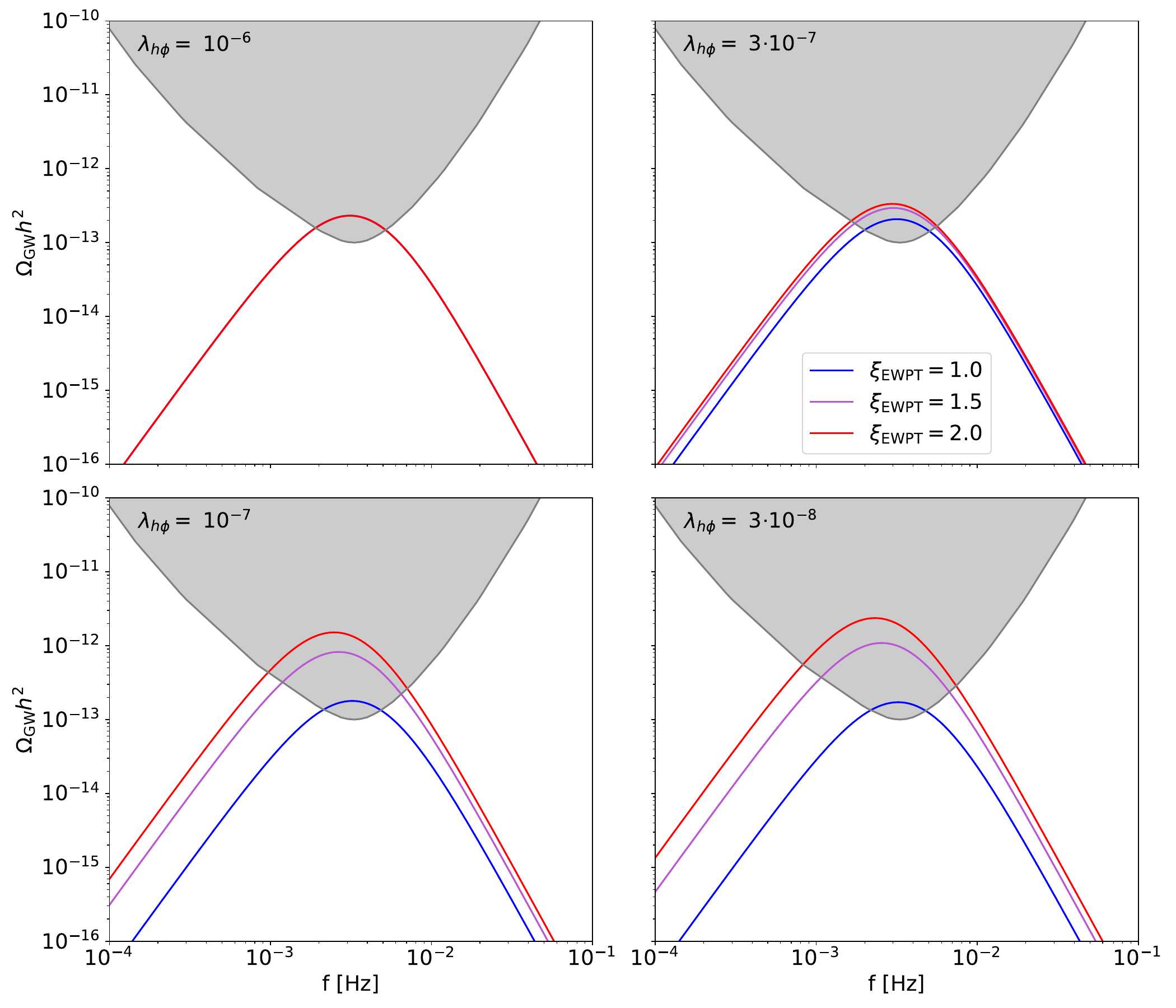}
  \caption{Gravitational wave signal for the same scenarios as considered in figure~\ref{fig:evolution},
  i.e.~for various values of the portal coupling $\lambda_{h\phi}$ (different panels) and the initial
  temperature ratio $\xi_\EWPT$ (different colours).
  }
  \vspace{-5mm}
  \label{fig:gw_with_xi}
\end{figure}

We are finally in the position to calculate the GW signal for $\xi \neq 1$ and
$\lambda_{h\phi} < 10^{-6}$. We show the result in figure~\ref{fig:gw_with_xi}, for various
initial values of the temperature ratio as well as portal couplings. As
anticipated, a net enhancement of the GW signal is found for $\xi_\EWPT > 1$, provided
that $\lambda_{h\phi}$ is sufficiently small for the sectors to not equilibrate before the
PT (cf.~figure~\ref{fig:evolution}). The enhancement saturates
for $\xi_\EWPT \gtrsim 2$, see also the discussion in ref.~\cite{Ertas:2021xeh}, 
implying a dark sector energy density that initially dominates over that of the SM sector.
For too small portal couplings $\lambda_{h\phi} \lesssim 3 \times 10^{-8}$, on the other hand,
the effect of dilution becomes relevant and the GW signal starts to become suppressed. Crucially,
changing $\xi$ and $\lambda_{h\phi}$ does not significantly affect the peak frequency,
such that the GW signal remains within the LISA sensitivity range.

\begin{figure}[t]
  \centering \includegraphics[width=0.95\textwidth]{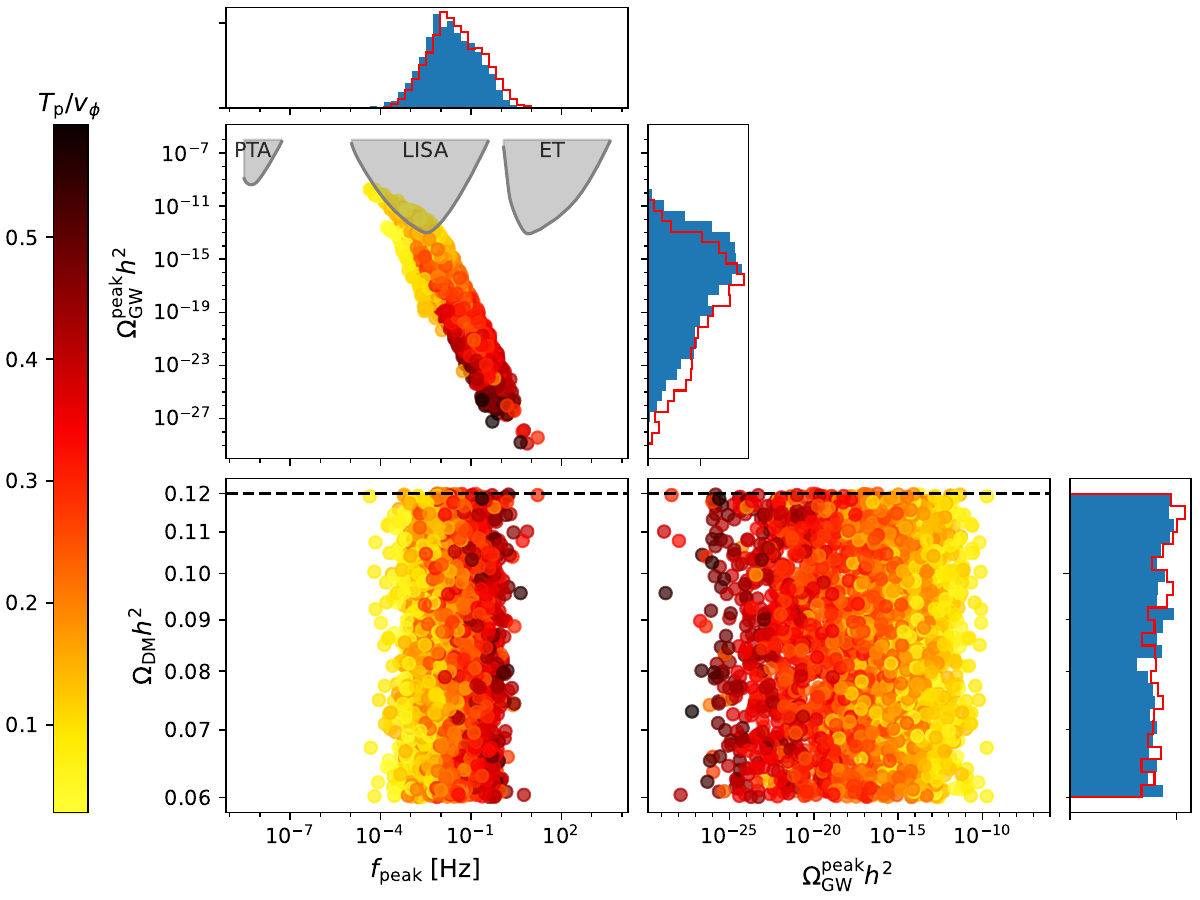}
  \caption{Same as figure~\ref{fig:triangle-restr}, but without the assumption of thermal
    equilibrium between the two sectors. Specifically, we consider $\xi_\EWPT = 2$ and
    $\lambda_{h\phi} = 10^{-7}$. Compared to the situation in figure~\ref{fig:triangle-restr} 
    (indicated by the red lines in the
    1D histograms), the GW amplitude is shifted
    to slightly larger values, while the peak position remains almost unaffected.}
  \label{fig:triangle-xi2}
\end{figure}

We can now test the robustness of our results from section~\ref{sec:relic}, where we
assumed $\xi = 1$ and $\lambda_{h\phi} > 10^{-6}$, by allowing larger values of $\xi$ and smaller
values of $\lambda_{h\phi}$. In figure~\ref{fig:triangle-xi2} we show the same result as in
figure~\ref{fig:triangle-restr}, but now for $\xi_\EWPT = 2$ and
$\lambda_{h\phi} = 10^{-7}$. For comparison we show the 1D distributions from
figure~\ref{fig:triangle-restr} as red lines. In this plot we have removed points
  for which the dark sector temperature still differs significantly from the SM
  temperature when the dark Higgs boson becomes non-relativistic, i.e.\ for which
  $\xi_\text{nr} > 1.1$ at $T_{\DS,\text{nr}} = m_\phi / 3$. The reason is that for such
  cases our final predictions depend on the details of chemical decoupling within the dark
  sector, which we do not study further in this work. While this requirement removes
  almost half of the parameter points, it does not introduce any significant bias, i.e.\
  the distributions of $f_\text{peak}$ and $\Omega_\text{GW}^\text{peak} h^2$ look very similar
  with and without this additional requirement.

The parameter combination $\xi_\EWPT = 2$ and
$\lambda_{h\phi} = 10^{-7}$ leads to a nearly maximal enhancement of the GW signal. As expected, we find that the peak position of the GW signal is not affected, such
that the frequency range implied by the observed DM relic abundance remains within the
LISA sensitivity window. The amplitude of the GW signal, on the other hand, is slightly
enhanced, as can be seen from the comparison in the corresponding one-dimensional
histogram.

We can make this statement more precise by once again interpreting the density of points in the scatter plots as a probability 
distribution for the observables. Compared to figure~\ref{fig:triangle-restr}, we find that the probability to obtain a signal 
observable with LISA increases from 3\% to 8\%. Limiting ourselves to parameter regions with strong supercooling, the fraction of observable events increases from 35\% to 69\%. We summarize our findings in table~\ref{tab:observable}.

\begin{table}[t]
 \centering
 \begin{tabular}{p{0.41\textwidth}C{0.25\textwidth}C{0.25\textwidth}}
 \toprule
  & \multicolumn{2}{c}{\centering Fraction of parameter points observable by LISA} \\
  & $\xi_\text{EWPT} \!=\! 1$, $\lambda_{h\phi} \!=\!  10^{-6}$ & $\xi_\text{EWPT} \!=\!  2$, $\lambda_{h\phi} \!=\!  10^{-7}$ \\
  \midrule
  Full sample & 0.1\% & 0.5\% \\[1ex]
  First-order PT & 0.8\% & 3\% \\
  First-order PT + relic density & 3\% & 8\% \\[1ex]
  Strong supercooling & 10\% & 21\% \\
  Strong supercooling + relic density & 35\% & 69\% \\
  \bottomrule
 \end{tabular}
 \caption{Fraction of parameter points that predict an observable GW signal for LISA after imposing various selection requirements on the sample of points drawn from the parameter ranges discussed in section~\ref{sec:scan}.}
 \label{tab:observable}
\end{table}

We emphasize that this large increase is a result of fixing $\xi$ and $\lambda_{h\phi}$ 
to particular values. If we instead vary
  $\xi$ and $\lambda_{h\phi}$ as part of the scan, most parameter combinations will either give very
  similar results to the case $\xi = 1$ considered in section~\ref{sec:relic} or lead to an
  extended period where the dark sector energy density dominates.

\section{Conclusions}
\label{sec:conclusions}

In this work we explored correlations between the DM relic density and GW signals
arising from a first-order PT that breaks a $U(1)'$ gauge symmetry and gives rise to the
mass of the fermionic DM particle. We demonstrated that, while the amplitude of the GW
signal depends on the details of the effective potential and can vary over many orders of
magnitude, the peak frequency is tightly constrained once we impose the observed value for
the DM relic abundance. Intriguingly, the peak frequency is found to lie exactly in the
milli-Hertz range, which will be explored by the LISA mission.

The dark sector considered in this work is characterised by four parameters: the dark
gauge coupling $g$, the quartic coupling of the dark Higgs field $\lambda$, the dark Yukawa
coupling $y$ and the dark Higgs vev $v_\phi$. As a first step, we calculated the effective
potential and the percolation temperature of the PT and identified the regions of
parameter space that give a strong (large $\alpha$) and not too fast
($\beta/H\sim10^2$--$10^4$) PT, corresponding to potentially observable GW signals. We showed in
particular that large GW signals require sizeable couplings $g$ and
$\lambda$ and occur also for large values of $y$.
The relic density of the dark sector is determined by annihilations of DM fermions into
pairs of dark Higgs bosons. The requirement to match the observed DM relic density then
requires that the DM fermion cannot be much lighter than the dark Higgs boson, with mass $m_\phi\sim v_\phi$, which in
turn implies a sizable Yukawa coupling and a DM mass that is comparable to the dark Higgs
vev $v_\phi$. The dark Higgs vev, on the other hand, determines the percolation temperature and
hence the peak frequency of the resulting GW signal. This connection leads to a tight
correlation between relic density and GW peak frequency. Through comprehensive scans of
the parameter space, we confirmed that this correlation is indeed highly generic in our
model.

A rigorous statistical interpretation of our results is beyond the scope of this work, but
some estimates of the significance of the samples can be performed. A rough measure of the
fine-tuning required to have a visible signal at LISA can be obtained by assuming that the
sampling distributions of the parameters act as prior probabilities, and that the sampling
density of points hence indicates the posterior distributions of derived quantities. Indeed, the majority of points drawn from these distributions do not feature any first-order PT at all. Out of the points that feature a strongly-supercooled PT ($T_\perc / T_\crit < 0.5$), the probability of producing a visible
signal at LISA in our model is around 10\%. With the additional requirement that the observed DM
relic abundance be reproduced, this probability increases to 35\%, as a result of the strong correlation between the predicted relic density and the peak frequency of the GW signal (see figure \ref{fig:triangle-restr}).

We then studied two connected questions: How does the dark sector transfer its energy
density to the Standard Model? And is it justified to assume the same temperature for both
sectors? Indeed, the PT leads to an increase of the dark sector temperature, as vacuum
energy is converted into rest mass and kinetic energy. Having confirmed that
the dark sector itself thermalises immediately after the PT, we discussed in detail how
the two sectors thermalise with each other for the specific case that the two Higgs bosons
in the theory interact via the portal coupling $\lambda_{h\phi}$. After both electroweak symmetry breaking and dark sector
symmetry breaking, this interaction leads to mixing between the Higgs bosons, such that
they can decay into each other as well as into fermions of both sectors.

We derived and solved the Boltzmann equations for the entropy transfer between the
two sectors and showed that for $\lambda_{h\phi} > 10^{-6}$ the assumption of equal temperature for
both sectors is well justified. This portal coupling is small enough to be consistent with
all laboratory constraints, in particular with direct detection experiments. For even
smaller portal couplings, on the other hand, the temperatures of the two sectors may differ
significantly, motivating us to consider the initial temperature ratio $\xi_\EWPT$ as a
free parameter.
While small values of $\lambda_{h\phi}$ combined with large initial values of $\xi_\EWPT$ may lead to
an increase in the amplitude of the GW signal, the dark Higgs bosons will decay only
slowly after the PT and may end up dominating the energy density of the universe. The
resulting entropy injection would then lead to a substantial dilution of GW signals. We find 
that for $\lambda_{h\phi} \approx 10^{-7}$ a net enhancement
remains, demonstrating that it is possible to have $\xi > 1$ during the dark sector PT while at the same
time avoiding the dilution of the GW signal due to sufficiently rapid thermalisation afterwards.
Rpeating our parameter scans for $\xi = 2$ and $\lambda_{h\phi} = 10^{-7}$, we 
find that the impact on the GW spectrum is
only modest; while the typical amplitude is slightly enhanced, the peak frequency remains unchanged. In combination, these effects increase the fraction of points with a strongly supercooled PT that would be observable in LISA to 69\%. Hence, our conclusion
regarding the correlation between the DM relic abundance and the GW peak frequency applies also to dark sectors that
thermalise only slowly with the Standard Model.

An interesting open question is how the relic density calculation would change for even smaller values of the portal coupling than
what we consider. In this case, the energy density of the universe would be dominated by non-relativistic dark Higgs bosons,
which may develop a chemical potential if number-changing processes are inefficient. The relic density calculation then requires
solving a coupled set of Boltzmann equations with non-trivial evolution of the dark sector temperature. While the details of this
calculation are beyond the scope of this work, the general expectation is that the DM relic abundance would be increased. This
might open up the possibility to have a dark sector PT in the nano-Hertz frequency range and hence of interest in the context of
recent results from pulsar timing arrays~\cite{NANOGrav:2023gor, EPTA:2023fyk, Reardon:2023gzh}.
Another promising avenue for further investigations opens up due to our findings regarding the bubble wall velocity presented in 
appendix~\ref{app:darkwalls}. There exists a potentially relevant part in the parameter space of our model in which the 
Bödeker-Moore criterion hints towards non-relativistic bubble wall velocities. For such low wall speeds the effect of bubble 
filtering~\cite{Baker2019} may be non-negligible, and the calculation of the DM relic abundance is expected to be more involved than presented here.

Let us finally mention that even for tiny portal couplings there is a chance to actually
detect the DM particles that we consider: unlike for annihilation into dark Higgs boson pairs, 
the mixed annihilation channel into one dark
Higgs boson and one dark photon is not suppressed in the limit of small DM velocities.
If kinematically allowed, it may thus lead to observable signals in indirect detection experiments~\cite{Bell:2016fqf} 
such as CTA~\cite{CTA:2020qlo}. 
Such an observation would raise the possibility to explore in
practice the correlations studied in this work and pin down the detailed structure of the
dark sector.

\acknowledgments

We thank Thomas Konstandin and Kai Schmidt-Hoberg for helpful comments on the manuscript and Peter Athron, Csaba Balazs, Andreas Ekstedt, Fatih Ertas, Henda Mansour, Lachlan Morris and Margarethe
M\"{u}hlleitner for discussions. CT thanks FNAL for its hospitality. This project has received funding from the European Union’s Horizon Europe research and innovation programme under the Marie Skłodowska-Curie Staff Exchange  grant agreement No 101086085 – ASYMMETRY, and from the Deutsche Forschungsgemeinschaft (DFG) through the Emmy Noether Grant No. KA 4662/1-2 and grant 396021762~--~TRR~257.

\newpage
\appendix
\section{Dark bubble walls}
\label{app:darkwalls}

In addition to the strength $\alpha$ of the PT, its speed $\beta/H$, the percolation
temperature $T_\perc$ and the temperature ratio $\xi_\perc$, it is also necessary to know the speed of the bubble
walls $\vw$ in order to determine the processes that dominate the GW
signal from a dark sector PT. While the former parameters can be obtained from the
effective potential $V_\text{eff}(\phi, T)$, the bubble wall velocity depends on plasma
effects of the expanding bubble walls and therefore requires additional considerations.
For bubbles expanding into the vacuum (i.e.\ if bubbles expand into a plasma that is not
influenced by a change of the scalar vev), there is no source of friction, such that
bubble walls can accelerate up to the point of their collision. For walls that interact
with the surrounding plasma, on the other hand, several model-dependent sources of
friction have been discussed~\cite{Liu:1992tn, Bodeker:2009qy, Bodeker:2017cim,
  Azatov:2020ufh, GarciaGarcia:2022yqb}. If the friction increases with the bubble wall
velocity, the acceleration of the bubble walls eventually stalls and a terminal velocity
is reached. In this case, the bulk motion of the plasma dominates the GW
spectrum~\cite{Azatov:2023xem}.

In this paper we take the approach suggested in ref.~\cite{Vanvlasselaer:2022fqf}, i.e.\
we show that $\vw$ is either expected to be non-relativistic in our model or that the
bubbles are relativistic, but do not run away due to the emission of soft gauge bosons in
the bubble
walls.
We conclude that the plasma motion is responsible for the dominant part of the GW signal
and the contribution from bubble wall collisions is negligible.

To decide whether a bubble wall can accelerate up to relativistic velocities, we use the
B\"{o}deker-Moore criterion~\cite{Bodeker:2009qy}, which relates the velocity-independent
leading-order (LO) bubble wall friction $\mathcal{P}_\mathrm{LO}$ to the amount of liberated vacuum
energy density $\Delta V_\text{eff}$:
\begin{align}
	\text{Bödeker-Moore criterion:} \quad \begin{cases}
		\Delta V_\text{eff} > \mathcal{P}_\mathrm{LO} \quad \text{Relativistic bubble walls} \\
		\Delta V_\text{eff} < \mathcal{P}_\mathrm{LO} \quad \text{Non-relativistic bubble walls} \label{eq:Bodeker}
	\end{cases} \,.
\end{align}
We emphasize that this criterion is insufficient to decide whether walls can run away
(i.e.~accelerate indefinitely), because of the next-to-leading-order (NLO) friction
$\mathcal{P}_\mathrm{NLO}$, which scales with powers of
$\gamma_\text{w} = 1/\sqrt{1 - \vw^2}$~\cite{Bodeker:2017cim,Hoche:2020ysm,Gouttenoire:2021kjv}. Bubbles can run away only if
$\Delta V_\text{eff} > \mathcal{P}_\mathrm{LO} + \mathcal{P}_\mathrm{NLO}$ for all $\vw$. Otherwise they will reach
a relativistic terminal velocity given by the equilibrium of forces,
$\Delta V_\text{eff} = \mathcal{P}_\mathrm{LO} + \mathcal{P}_\mathrm{NLO}$.

The LO friction due to particles acquiring a mass when crossing the bubble wall is given
by~\cite{Bodeker:2009qy, Espinosa:2010hh}
\begin{align}
	\mathcal{P}_\mathrm{LO} \simeq \sum_i g_i \, c_i \frac{\Delta m_i^2}{24} T_\nuc^2 \,, \label{eq:LOfriction}
\end{align}
where $\Delta m_i^2$ is the positive change of the mass square of particle species $i$ during
the PT, $g_i$ is the corresponding number of degrees of freedom, and $c_i = 1$ $(1/2)$ for
bosons (fermions). This expression assumes that the particle masses outside the bubble are
below the dark sector temperature. For ultra-relativistic bubble walls, the production of
heavy particles with mass up to $\gamma_\text{w} T_\nuc$ can also add to the LO
friction~\cite{Azatov:2020ufh}. For the dark sector considered in this work, all particles
are light before the PT and subsequently obtain a mass comparable to the scale of the PT,
such that \eqref{eq:LOfriction} is the only relevant contribution to the LO friction.

\begin{figure}
	\centering
	\includegraphics[width=\linewidth]{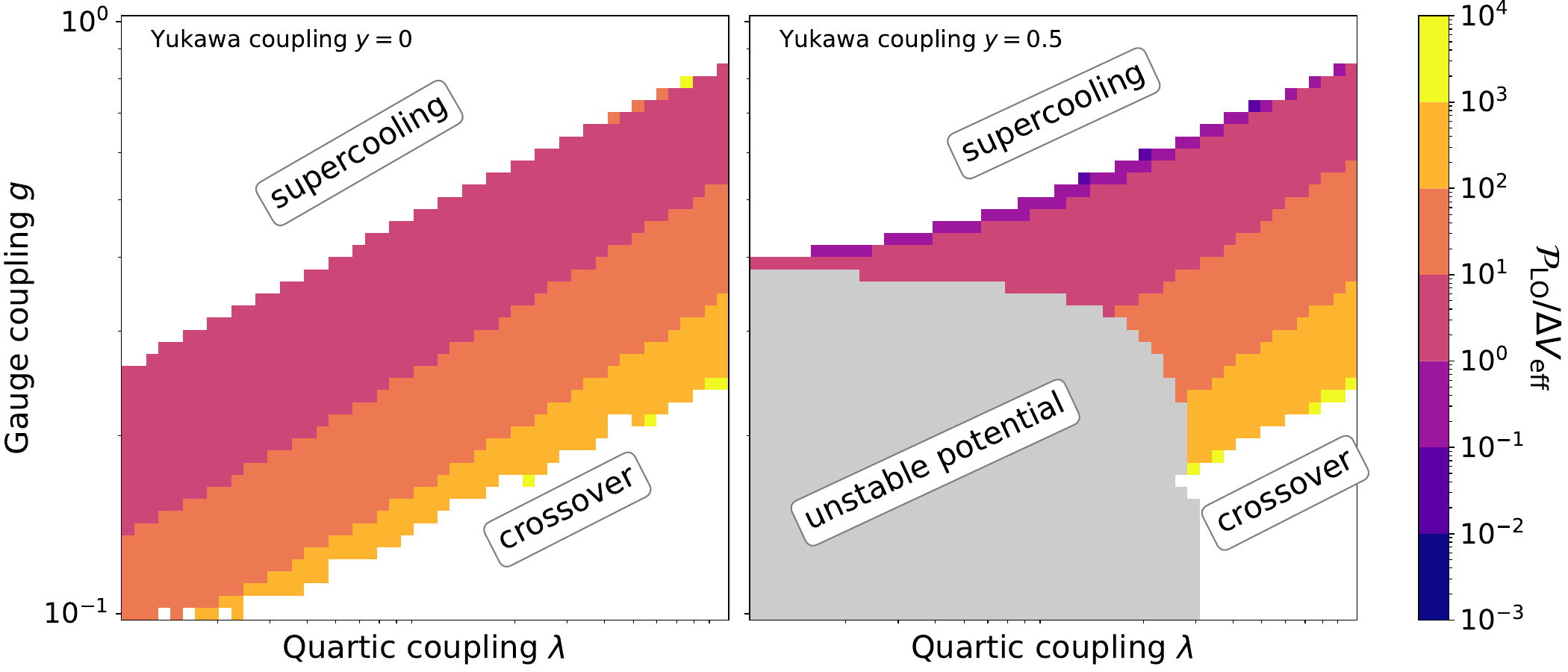}
	\caption{The leading-order friction $\mathcal{P}_\text{LO}$ over the difference
          $\Delta V_\text{eff}$ in potential energy between the true and false vacuum phases as
          a function of the quartic coupling $\lambda$ and the gauge coupling $g$ for values of
          the Yukawa coupling $y = 0$ and $y = 0.5$. Values greater than one indicate that
          relativistic bubble wall velocities cannot be reached,
          cf.~eq.~\eqref{eq:Bodeker}. For values smaller than one, a relativistic terminal
          velocity is expected.}
	\label{fig:verifyvw}
\end{figure}

In the case of a broken $U(1)'$ with a fermionic species, the LO friction
reads~\cite{Espinosa:2010hh}
\begin{align}
	\mathcal{P}_\text{LO} \simeq \left(3 m_{A'}^2 + m_\phi^2 + 2 m_\chi^2\right) \, \frac{{T_{\perc}}^2}{24} \approx \frac{m_{A'}^2 \, {T_{\perc}}^2}{8} \,,
\end{align}
where in the last step we have used that the dominant contribution comes from the heaviest
state in the dark sector, which in the parameter regions of interest is the dark photon.
The amount of released vacuum energy $\Delta V_\text{eff}$ can be estimated from the
zero-temperature potential, which gives
\begin{equation}
\Delta V_\text{eff} \approx \frac{\lambda}{4} v_\phi^4 = \frac{m_\phi^2}{8} v_\phi^2 \,.
\end{equation}
Hence we find that the B\"{o}deker-Moore criterion for relativistic bubble walls is
satisfied if $m_\phi / m_{A'} > {T_{\perc}} / v_\phi$, which is the case for the parameter
regions that give strongly supercooled PTs corresponding to observable GW signals
($T_{\perc} \lesssim 0.1 \, v_\phi$). In these parameter regions, the bubble walls are therefore
expected to be relativistic, $\vw \rightarrow 1$. This finding also implies that we can neglect the
effect of bubble filtering, which is only relevant in the (deeply) non-relativistic regime
$\vw \ll 1$~\cite{Baker2019,Chway2019}. In the regions in which weaker GW signals are
expected, the Bödeker-Moore criterion instead hints towards slower bubble walls, see
figure~\ref{fig:verifyvw}.

The NLO friction created by the emission of soft dark photons into the broken phase
quickly starts to grow with $\gamma_\text{w}$. The bubble walls will therefore reach a
terminal, asymptotic bubble wall velocity which is close to the speed of light. The
precise value of $\gamma_\text{w}^\text{terminal}$ is unnecessary for our purposes, as the
existence of a terminal yet relativistic bubble wall velocity is sufficient to assume a
dominant GW emission through bulk fluid motion. A more refined calculation of the
respective energy budgets for the processes emitting gravitational radiation was performed
in refs.~\cite{Lewicki:2022pdb,Ellis:2020nnr, Ellis:2019oqb}. In
ref.~\cite{Lewicki:2022pdb} it was shown that for sufficiently high terminal bubble wall
velocities the fluid profiles are strongly peaked, such that the emitted GW spectral
shapes are in fact indistinguishable from bubble collisions. We conclude that it is hence
a reasonable approximation to work with a GW spectrum that is solely sourced through sound
waves.

\section{Boltzmann equations for thermalisation and dark matter freeze-out}
\label{app:thermalisation}

Here we discuss the processes we consider for the entropy transfer between the
DS and the SM.

\subsection{Thermal mixing angle}

The Higgs mixing angle defined in eq.~\eqref{eq:mixing-angle} depends on temperature
through the masses and vevs of the two Higgs bosons. The temperature dependence of the
dark Higgs boson can be directly obtained from the effective potential, whereas we follow
ref.~\cite{Bringmann:2021sth} to implement the temperature dependence of the SM Higgs
boson. For large values of the dark Higgs vev, we sometimes encounter the situation that
the SM Higgs and dark Higgs mass become similar or even cross, in which case the mixing
angle apparently diverges. To regulate this unphysical divergence, we have to include the
finite width $\Gamma_{h}$ of the Higgs resonance. As shown in ref.~\cite{Fradette:2018hhl},
including the width leads to an effective mixing angle given by
\begin{align}
  \theta_{\mathrm{eff}}^2(T) = \frac{\left(\lambda_{p}v_{h}(T)v_{\phi}(T) \right)^2}{(m_{\phi}^2(T) -
  m_{h}^2(T))^2 + (m_{\phi}(T)\Gamma_{h})^2} \,.
\end{align}

\subsection{The Boltzmann equation for entropy transfer}

In our analysis we specify the initial conditions, i.e.\ the temperature ratio $\xi_\EWPT$ of the
dark sector and SM bath at the EWPT, for which we take $T_\EWPT = 150\,$GeV. Which
processes contribute to the thermalisation of the two sectors depends on whether or not
the $U(1)'$ gauge symmetry is already broken at this point. We consider two different
timelines:
\begin{itemize}
\item The EW symmetry breaks before the dark sector PT. This is the case for the majority
  of our parameter space. Here, the thermalisation between the two sectors is initially
  determined by the processes $h h \leftrightarrow \Phi\Phi$ and the decay of the SM Higgs
  $h \rightarrow \Phi\Phi$. Additional processes can only contribute after the dark sector PT.
\item The dark sector PT occurs before EW symmetry breaking. This can happen for parameter points with a
  large vev of the dark Higgs and not too strong supercooling. Once both symmetries are broken, thermalisation proceeds via $hh \leftrightarrow \phi\phi$, the decays of the SM Higgs into dark Higgs
$h \rightarrow \phi\phi$ if it is kinematically allowed, and the decays of the dark Higgs into SM
particles $\phi \rightarrow \mathrm{SM} \ \mathrm{SM}$ through Higgs mixing.
\end{itemize}
In the following we give the relevant
expressions contributing to the entropy transfer.

\subsubsection{$2\to 2$ processes}
\label{subsec:2to2processes}

In most regions of parameter space, the dark sector phase is unbroken immediately after the EWPT, such that there are no dark Higgs boson decays that can transfer entropy between the two sectors. Here the
$2 \to 2 $ process induced by the portal coupling $\Phi\Phi \leftrightarrow hh$ become relevant. Since the
particles' thermal masses are smaller than the temperature, we cannot take the usual
non-relativistic approach. Instead we will follow the relativistic treatment of the
calculation of the entropy transfer developed in refs.~\cite{arcadi2019,deromeri2020}, which we
briefly sketch here. The heat transfer for $2\to2$ processes can be expressed as
\begin{align}\label{eq:entropy-22}
  \dot{q}\big|_{2\to2} &=  \int \frac{\mathrm{d}^3 p}{(2\pi)^3} E C[f] \nonumber\\
  &= \int \frac{\dd^3p_1}{(2\pi)^{3}}\frac{\dd^3p_2}{(2\pi)^{3}2E_{2}}\frac{\dd^3k_1}{(2\pi)^{3}2E_{k_{1}}}
    \frac{\dd^3k_2}{(2\pi)^{3}2E_{k_{2}}} |\mathcal{M}|^2 (2\pi)^4\delta(\Sigma p) \nonumber \\
  & ~~~~~~~~~~~~~~~~~~~~~~~~~~~~~~~~~~~~~~~~\times f(p_{1})f(p_2)(1+ f(k_{1}) (1+f(k_2)) \nonumber\\
    &= \int \frac{\dd^3p_1}{(2\pi)^{3}2E_{1}}\frac{\dd^3p_2}{(2\pi)^{3}2E_{2}} 8 E_{1} F(p_1, p_2) \sigma(p_1, p_2) f(p_1) f(p_2)\,.
\end{align}
Here the final state statistical factors are absorbed into the cross section and
$F(p_{1}, p_{2}) = \sqrt{(p_1\cdot p_2)^2-m^2_{1}m_2^2}$. It is easiest to calculate this
cross section in the center of mass frame. However, since the Bose-Einstein and
Fermi-Dirac distributions are not Lorentz-invariant we have to apply a Lorentz boost
$\Lambda$ from the cosmic rest frame where $u = (1, 0, 0, 0)^{T}$ into the center of mass frame,
which can be parameterised by the rapidity $\eta$ and two angles $\theta$ and $\varphi$; for details see
ref.~\cite{arcadi2019}. The phase space distribution becomes
\begin{align}
  f(k) = \frac{1}{e^{u \cdot k/T} \mp 1} \overset{\Lambda}{\rightarrow} f_{\Lambda}(k) =
  \frac{1}{e^{\left( k_{0} \cosh\eta + k_{1}\sinh\eta  \right)/T }\mp 1}\,.
\end{align}
With this, we can rewrite the center of mass cross section as
\begin{align}
  \sigma_{\mathrm{CM}}(p_1,p_2) = \frac{1}{(2\pi)^216F(p_1,p_2)} \int \dd \varphi \dd \cos\theta \left| \mathcal{M} \right|^2
  \frac{\sqrt{E^2-m_{f}^2}}{E} \left( (1+f_{\Lambda}(k_1))(1+ f_{\Lambda}(k_2)) \right)\,.
\end{align}
The matrix element for the processes $\Phi\Phi\to hh$ is at tree level simply given by
$\mathcal{M} = i\lambda_{h\phi}$. For this angle-independent transition amplitude, we can integrate over the
solid angle and obtain
\begin{align}
  \sigma_{\mathrm{CM}} = \frac{|\mathcal{M}|^2}{64\pi E^2} \frac{T}{\sqrt{E^2-m_{\Phi}^2} \sinh{\eta}}
  \frac{1}{1-e^{- \frac{2E}{T}\cosh\eta}}  \ln \underbrace{\left[ \frac{\sinh \left( \frac{E\cosh\eta +
  \sqrt{E^2-m_{h}^2}\sinh\eta}{2T} \right)}{\sinh \left( \frac{E\cosh\eta -
  \sqrt{E^2-m_{h}^2}\sinh\eta}{2T} \right)} \right]}_{\equiv \lambda(E, \eta, T, m_{h})} \, .
\end{align}
In the case of initial and final states with respectively equal masses,
eq.~\eqref{eq:entropy-22} reduces to
\begin{align}
  I_{\Phi\Phi \rightarrow h h}  \equiv \frac{2T_{\mathrm{DS}}}{\pi^4} \int_{m_{\Phi}(T)}^{\infty}\dd E \sqrt{E^2-m_{\Phi}^2}E^4
  \int_{0}^{\infty}\dd \eta \frac{\sinh\eta\cosh\eta}{e^{2E\cosh\eta/T_{\mathrm{DS}}}-1}
  \ln \lambda(E, \eta, T_\mathrm{DS}, m_{\Phi}) ~ \sigma_{\mathrm{CM}} \, .
\end{align}
This expression can now be efficiently evalutated numerically. An analogous expression is
obtained for the process $h h \rightarrow \phi \phi$.

We note that in principle there are additional $2 \to 2$ processes that may contribute to thermalisation. The process $\phi \phi \to t \bar{t}$ via an off-shell SM Higgs boson is strongly Boltzmann-suppressed below the EWPT~\cite{Bringmann:2021sth} and does not give a relevant contribution in the temperature range that we consider. However, processes such as $\phi + q \to g + q$ with a quark in the $t$-channel can give a relevant contribution if the decay $\phi \to b \bar{b}$ is kinematically forbidden~\cite{Evans:2017kti,Fradette:2018hhl}. Since this is the case only in a very small fraction of the parameter space that we consider, we neglect these processes, thus giving a conservative estimate of the thermalisation rate.

\subsubsection{Standard Model Higgs boson decays}
\label{app:inverse-SM-Higgs-decays}

After both symmetries are broken and for temperatures comparable to the SM Higgs boson
mass, a second process of interest besides dark Higgs decays is the resonantly enhanced
pair-annihilation of dark Higgs bosons into predominantly bottom quarks:
$\phi \phi \to h \to b \bar{b}$. In thermal equilibrium, the rate of this process can be related
to the inverse process, which is the decay $h \to \phi \phi$ with partial width given by
\begin{equation}
  \Gamma_{h\to\phi\phi} = \frac{(m_h^2 + 2 \, m_\phi^2)^2 \, \sin^2 2 \theta_{\mathrm{eff}}}{128 \pi \, m_h}
  \left(1 - \frac{4 \, m_\phi^2}{m_h^2}\right)^{1/2} \left(\frac{1}{v_\phi} \cos \theta_{\mathrm{eff}}
    + \frac{1}{v_h} \sin \theta_{\mathrm{eff}} \right)^2 \,.
\end{equation}
This gives the additional term in the heat transfer rate
\begin{align}
  \dot{q}\big|_{h\to\phi\phi} =  - m_{h}\Gamma_{h \to \phi \phi} n^\text{eq}_h
  \left[ 1 - \left(\frac{n_\phi}{n^\text{eq}_\phi}\right)^2 \right]\,.
\end{align}
Before the dark sector symmetry breaking, we also have the decay $h \rightarrow \Phi\Phi$ with the decay rate
\begin{align}
   \Gamma_{h\to\Phi\Phi} = \frac{\lambda_{h\phi}^2 v_{h}^{2}}{128 \pi \, m_h} \,,
\end{align}
which can be treated in analogy to the case above.

\subsubsection{Full Boltzmann equation}

The full Boltzmann equation that we solve before the PT, in case it occurs after the
EWPT then reads
\begin{align}
  \dot{s}_{\mathrm{DS}} + 3H s_{\mathrm{DS}} = - \frac{m_h}{T_{\mathrm{DS}}}\Gamma_{h\rightarrow\Phi\Phi} n_{h}^{\mathrm{eq}}
  \left[\left(\frac{n_{\phi}}{n_{\phi}^{\mathrm{eq}}}  \right)^2 - 1  \right]- \frac{1}{T_{\DS}} I_{\Phi\Phi\to hh}
  + \frac{1}{T_{\DS}}I_{hh\to \Phi\Phi} \,. 
\end{align}
The equation for the case of the PT occuring before the EWPT follows
analogously. After both symmetries are broken the full Boltzmann equation reads
\begin{align}
  \dot{s}_{\mathrm{DS}} + 3H s_{\mathrm{DS}}
  &= - \frac{m_\phi}{T_\mathrm{DS}}\Gamma_{\phi \to \mathrm{SM}\mathrm{SM}} n^\text{eq}_\phi
  \left( \frac{n_\phi}{n^\text{eq}_\phi} - 1 \right) + \frac{m_{h}}{T_\mathrm{DS}}\Gamma_{h \to \phi \phi} n^\text{eq}_h
  \left[ 1 - \left(\frac{n_\phi}{n^\text{eq}_\phi}\right)^2 \right] \notag\\
  &- \frac{1}{T_{\DS}} I_{\phi\phi\to hh} + \frac{1}{T_{\DS}}I_{hh\to \phi\phi} \,.
\end{align}
where we use the decay widths of the SM Higgs into other SM particle pairs from ref.~\cite{Ferber:2023iso}. The equations for the SM entropy follow analogously.

\subsection{Annihilation cross sections}
\label{app:annihilation}

In the following we list the various DM annihilation cross sections in the non-relativistic limit, up to second order in the
CMS velocity $v$ (of each of the DM particles):
\begin{align}\label{eq:sig-cc-aa}
  (\sigma v)_{\chi\chi \to A^{\prime}A^{\prime}}
  = &   \frac{m_{A'}^4 (m_\chi^2 - m_{A'}^2)^{3/2}}{64 \pi \, v_\phi^4 \, m_\chi(m_{A'}^2 - 2 m_\chi^2)^2} \nonumber \\ & +
     \frac{v^2 \sqrt{m_\chi^2 - m_{A'}^2}}{384 \pi \, v_\phi^4 \, m_\chi (m_\phi^2 - 4 m_\chi^2)^2 (m_{A'}^2 - 2 m_\chi^2)^4} \nonumber \\ & \quad \times
     \Big[144 m_{A'}^{12} m_\chi^2  
     + 2 m_{A'}^{10} (7 m_\phi^4 - 88 m_\phi^2 m_\chi^2 - 432 m_\chi^4)  \nonumber \\ & \qquad 
     + 128 m_\chi^{10} (m_\phi^4 + 8 m_\chi^4) -
     64 m_{A'}^2 m_\chi^8 (3 m_\phi^4 + 16 m_\phi^2 m_\chi^2 +
     32 m_\chi^4) \nonumber \\ & \qquad 
     + 4 m_{A'}^4 m_\chi^6 (17 m_\phi^4 + 600 m_\phi^2 m_\chi^2 + 592 m_\chi^4) \nonumber \\ & \qquad +
     m_{A'}^8 (-73 m_\phi^4 m_\chi^2 + 1128 m_\phi^2 m_\chi^4 + 1840 m_\chi^6) \nonumber \\ & \qquad + 4 m_{A'}^6 (25 m_\phi^4 m_\chi^4 -
     648 m_\phi^2 m_\chi^6 - 496 m_\chi^8)\Big]\nonumber \\
  &+ \mathcal{O}\left(v^4\right)
  \end{align}
  \begin{align}
  (\sigma v)_{\chi\chi \to \phi\phi}
  = & \frac{v^2 m_\chi \sqrt{m_\chi^2 - m_\phi^2}}{192 \pi \, v_\phi^4 (m_\phi^2 - 4 m_\chi^2)^2 (m_\phi^2 - 2 m_\chi^2)^4} \left(3 m_\phi^4 - 8 m_\phi^2 m_\chi^2 + 8 m_\chi^4\right) \nonumber \\ & \quad \times  \left(9m_\phi^8 - 64 m_\phi^6 m_\chi^2 + 200 m_\phi^4 m_\chi^4 - 352 m_\phi^2 m_\chi^6 + 288 m_\chi^8\right) \nonumber\\
  & + \mathcal{O}\left(v^4\right) \\[2ex]
  (\sigma v)_{\chi\chi\to A^{\prime}\phi}
  = & \frac{(m_{A'}^4 - 2m_{A'}^2m_\phi^2 + m_\phi^4 -
     8m_{A'}^2m_\chi^2 -
     8m_\phi^2m_\chi^2 + 16m_\chi^4)^{3/2}}{2048 \pi v^4 m_\chi^4} \nonumber \\ & + \frac{v^2 \sqrt{m_{A'}^4 + (m_\phi^2 - 4m_\chi^2)^2 -
     2m_{A'}^2(m_\phi^2 + 4m_\chi^2)}}{12288 \pi \, v_\phi^4 m_\chi^4 (m_{A'}^2 - 4m_\chi^2)^2  (m_{A'}^2 + m_\phi^2 - 4m_\chi^2)^4} \nonumber \\ & \quad \times
     \Big[ - 2m_{A'}^{14}(11m_\phi^2 - 228m_\chi^2) 
           - 16m_\chi^4(m_\phi^2 - 4m_\chi^2)^4
             (15m_\phi^4 - 80m_\phi^2m_\chi^2 +
               16m_\chi^4) \nonumber \\ & \qquad
            -11m_{A'}^{16} - m_{A'}^{12}(-11m_\phi^4 -
               1200m_\phi^2m_\chi^2 + 7264m_\chi^4)
               \nonumber \\ & \qquad +
            4m_{A'}^{10}(11m_\phi^6 +
               266m_\phi^4m_\chi^2 -
               4328m_\phi^2m_\chi^4 + 15136m_\chi^6) \nonumber \\ & \qquad +
            8m_{A'}^2m_\chi^2
             (m_\phi^2 - 4m_\chi^2)^2
             (15m_\phi^8 - 244m_\phi^6m_\chi^2 +
               1280m_\phi^4m_\chi^4 -
               2880m_\phi^2m_\chi^6 + 6912m_\chi^8) \nonumber \\ & \qquad -
            m_{A'}^4(m_\phi^2 - 4m_\chi^2)^2
             (11m_\phi^8 - 344m_\phi^6m_\chi^2 +
               1808m_\phi^4m_\chi^4 -
               8704m_\phi^2m_\chi^6 + 80384m_\chi^8) \nonumber \\ & \qquad  -
            m_{A'}^8(-11m_\phi^8 -
               32m_\phi^6m_\chi^2 +
               12464m_\phi^4m_\chi^4 -
               114688m_\phi^2m_\chi^6 + 291840m_\chi^8) \nonumber \\ & \qquad  -
            2m_{A'}^6(11m_\phi^{10} -
               140m_\phi^8m_\chi^2 +
               1216m_\phi^6m_\chi^4 -
               29056m_\phi^4m_\chi^6 +
               201984m_\phi^2m_\chi^8 - 412672m_\chi^{10})\Big]\nonumber \\ &
    + \mathcal{O}\left(v^4\right)
\end{align}

\providecommand{\href}[2]{#2}\begingroup\raggedright\endgroup

\end{document}